\documentclass[12pt,a4paper]{article}

	\usepackage[english]{babel}
	\usepackage[latin1, utf8]{inputenc}
	\usepackage[T1]{fontenc}
	\usepackage{ae, aecompl, eurosym, bbold, dsfont, textcomp, lmodern, microtype}	

	\usepackage{amsmath, amssymb, amsthm}

	\usepackage{fancyhdr}		
	\usepackage{setspace}		
	\usepackage{titlesec}
		
	\usepackage{geometry}
	\usepackage{booktabs, adjustbox, float, rotfloat, rotating}
	\usepackage{caption, subcaption, enumitem}
	\usepackage{bookmark, pdflscape, ifthen, xurl}
	\usepackage{xcolor, pifont}		
 
	\usepackage{color}
		\definecolor{dark-red}{rgb}{0.6,0,0}
		\definecolor{listinggray}{gray}{0.9}
		\definecolor{darkgreen}{rgb}{0,0.4,0}
		\definecolor{lbcolor}{rgb}{0.9,0.9,0.9}	
	
\usepackage{comment}

	\usepackage{pgf,tikz}
	\usetikzlibrary{shapes.geometric, arrows}
	\tikzstyle{action} 	= [rectangle, rounded corners, minimum width= \textwidth*0.45, minimum height=2cm, text width=\textwidth*0.4, draw=black, fill=gray!5]	
	\tikzstyle{dataset} = [rectangle, minimum width= \textwidth*0.45, minimum height=2cm, text width=\textwidth*0.4, draw=black, fill=gray!25]
	\tikzstyle{arrow} = [thick,->,>=stealth]	
	
\tikzstyle{ConFrame} = [rectangle, draw=black, fill=gray!25]
\tikzstyle{Nature} = [circle, draw=gray!20, fill=gray!10]
\tikzstyle{Tier} = [rectangle, minimum width= \textwidth*0.15, minimum height=1cm, text width=\textwidth*0.05, draw=none, fill=gray!0]

	\usepackage{blkarray, multirow, multicol}
	\usepackage{tabularx, longtable, supertabular}
	\usepackage{ltcaption, enumerate}	
	
    \usepackage{arydshln}
	\usepackage{natbib}
	\usepackage{har2nat}

	\usepackage{hyperref}	
		\hypersetup{colorlinks=true,citecolor=blue}	

    \usepackage{graphicx}

	\usepackage[hang,flushmargin,multiple]{footmisc}

	\usepackage{titling}
\newcommand*\colourcheck[1]{%
  \expandafter\newcommand\csname #1check\endcsname{\textcolor{#1}{\ding{52}}}%
}
\colourcheck{green}
\colourcheck{yellow}
\colourcheck{red}

\newcommand*\colourx[1]{%
  \expandafter\newcommand\csname #1x\endcsname{\textcolor{#1}{\ding{56}}}%
}
\colourx{green}
\colourx{yellow}
\colourx{red}

\usepackage{etoolbox}
\AtBeginEnvironment{quote}{\par\singlespacing\small}
\makeatletter
\AtBeginEnvironment{quote}{\appto\@listi{\parsep\z@ \topsep\z@ \itemsep\z@}}

\begin{document}

\date{  \today }

\title{
 Good Guys With Guns? The Relationship Between Legal Firearm Ownership and Firearm Deaths and Crime in Canada
 \thanks{
 	\textbf{
 	We would like to thank Abel Brodeur, Louis-Philippe Beland, 
 	Aaron Chalfin, Mel Cross, Jamein Cunningham, Daniel Dutton, 
 	Rob Gillezeau, David B. Johnson, Lynda Khalaf, Caillin Langmann,
 	Cherie Metcalf, Thomas Russell, Jorge Tamayo and others for comments. 
 	We are grateful for discussions at the 
 	Canadian Economics Association 2022 Conference, 
        the Atlantic Canadian Economics Association 2022 Conference, 
 	Canadian Law and Economics 2022 Conference, and the
        2023 SCHNEPEL at Western Economics Association International Annual Conference. 
 	Special thanks to Noah Schwartz for guidance on the
 	Canadian gun ownership landscape, and Linoy Gofman for valuable research assistance.  
 	All errors are our own. Webb's research has been supported by grants from the Social Sciences
    and Humanities Research Council.}
	} 	
 }%

\author{Derek Mikola\thanks{Corresponding author: derek.mikola@gmail.com,
University of Ottawa \& Institute for Replication, Ottawa, ON.   } \\ University 
of Ottawa \& \\ Institute for Replication
	  \and Matthew D. Webb\thanks{\texttt{matt.webb@carleton.ca}} \\Carleton
	University } 

\maketitle

\onehalfspacing

\begin{abstract}
Civilian firearm ownership is politically contentious, yet most evidence linking guns to 
crime and death comes from the U.S. Canada offers a unique case: strict national regulations 
alongside widespread legal ownership. Using new administrative data from 2013-2019, we examine
associations between gun prevalence and homicides, suicides, and firearms-related crime. During 
this period, we find that handgun ownership rose 50\% and restricted licenses by 70\%. Recent 
legislation proposes a handgun freeze and rifle buyback. We assess whether these measures are 
likely to improve safety or whether existing laws already strike a balance. We find little
evidence of a strong relationship.  

\end{abstract}

\vfill


\vspace{6pt}


\vspace{12pt}

\clearpage

\section{Introduction} \label{section:intro}

%%%%%%%%%%%%%%%%%%%%%%%%%%%%%%%%%%%%%%%%%%%%%%%%%%%%%%%%%%%%%%%%%%%%%%
%Paragraph 1: Motivation
%%%%%%%%%%%%%%%%%%%%%%%%%%%%%%%%%%%%%%%%%%%%%%%%%%%%%%%%%%%%%%%%%%%%%%

Crime, homicides, and suicides are perpetually linked to 
firearms. In Canada during 2021, firearms were the primary cause of death in 
297 homicides (0.78 per 100,000) and were present in violent crime affecting
over 8,047 victims (27.4 per 100,000) \citep{Cotter2022}. At the same time,
it was estimated that Canadians had 34.7 civilian firearms per 100 people 
in 2017 (estimated around 12,708,000 total civilian firearms), fifth highest 
in the world.%
\footnote{Canada is behind the United States (120.5 per 100
people), Yemen (52.8 per 100 people), Montenegro (39.1 per 100 people), 
and Serbia (39.1 per 100 people) \citep{SAS2020}.} %
The United States has a similarly high relationship between firearms and related 
harms: over 56.5\% of suicides (27,593 deaths) and more than 76.2\% of homicides 
(15,364 deaths) involved firearms in the United States during 2024.%
\footnote{%
For suicides, see: \url{https://www.cdc.gov/nchs/fastats/suicide.htm}. %
For homicides, see: \url{https://www.cdc.gov/nchs/fastats/homicide.htm}. %
} %

%%%%%%%%%%%%%%%%%%%%%%%%%%%%%%%%%%%%%%%%%%%%%%%%%%%%%%%%%%%%%%%%%%%%%%
Arguments for how and why access to firearms may promote or deter 
violence are found in abundance. Within theoretical models, 
linking firearms and various outcomes like homicides and suicides can be
ambiguous. Giving credence to some theories and arguments for real-world 
decision making therefore relies on researchers' abilities to 
understand potential underlying relationships. Yet, the dearth of 
reliable data on firearms and gun owners leaves the 
research in this area wanting. 

%%%%%%%%%%%%%%%%%%%%%%%%%%%%%%%%%%%%%%%%%%%%%%%%%%%%%%%%%%%%%%%%%%%%%%
%Canada recent policy landscape 
Canadian regulations are a prime example of the importance of empirical evidence. 
Following the tragic Nova Scotia attacks, in April 2020, in which 22 people were killed, 
the federal government banned over 1,500 `assault-style' rifles \citep{Allen2022}. 
In 2022, the Canadian federal government introduced new firearms
legislation (Bill C-21, \textit{An Act to amend certain Acts and to make certain 
consequential amendments (firearms))} imposing many restrictions on 
licensed firearm owners.\footnote{More details can be found here \url{https://pm.gc.ca/en/news/news-releases/2022/05/30/further-strengthening-our-gun-control-laws}.}
The bill proposed both  (i) a ``freeze'' on the market for (legal) handguns, and (ii)
changing the classification of many long-guns from `unrestricted' or `restricted' to
`prohibited' and mandating a buyback.  The former precludes handguns from being bought,
sold, transferred or imported in Canada; effectively freezing the total number of handguns at
2022 levels. The latter would see assault-style firearms bought back at 
values (in Canadian dollars) between \$1,139 (VZ58 and variants) and 
\$6,209 (SG550 and SG551 and variants), with AR platforms (M16, AR-10, 
AR-15, M4 carbine and variants) and M14 rifles (and variants) being bought 
back at \$1,337 and \$2,612, respectively.  In December 2023, the bill received royal 
assent and the ``freeze'' on the market for (legal) handguns came into 
effect.\footnote{\url{https://www.publicsafety.gc.ca/cnt/cntrng-crm/frrms/c21-en.aspx}}
However, while the bill changed the classification of the affected long guns, as of July 
2025 the buyback for individuals has not yet been implemented.\footnote{See
\url{https://www.canada.ca/en/public-safety-canada/campaigns/firearms-buyback.html} 
for more details.}

%%%%%%%%%%%%%%%%%%%%%%%%%%%%%%%%%%%%%%%%%%%%%%%%%%%%%%%%%%%%%%%%%%%%%%
%Canada recent policy landscape 
Legislation change warrants questions of efficacy 
if enacted; specifically on how it will impact public safety through
the legal firearms channel. If legal firearms owners are the culprits 
of firearms-related crimes, then restricting access to firearms and 
the types of firearms available could be appropriate. 
But this need not be the case, and the relationship between legal
firearms and violent crimes remains unanswered. \citet[p.~5]{Allen2022} 
notes the current data gaps related to firearms in Canada: %
\begin{quote}%
\begin{singlespace}
\textit{It is important to recognize that there are limitations in our 
knowledge about firearms used in crime. There is little information 
currently collected about the characteristics of these firearms, 
such as details about the exact type of gun, who owned it (e.g., 
accused, victim, or someone else), how it was stored, or whether 
the owner was licensed.}

\medskip

\textit{Of particular concern, there is currently little information available 
to determine the source of firearms used in crime: for example, whether a 
gun used in a crime was stolen, illegally purchased or smuggled into the 
country. This information is sometimes not recorded by police services, 
recorded inconsistently or, in some cases, the information is simply not
available. For example, not all firearms are recovered from firearm-related 
homicides; consequently, only some of these guns are available for tracing.}
\end{singlespace}
\end{quote}%
And while police considered 46\% of firearm-related homicides to
be gang-related \cite[p.~3]{Moreau2022}, it is still uncertain exactly
to what extent organized crime is involved with firearm-related violence
\citep[p.~5]{Allen2022}.

%%%%%%%%%%%%%%%%%%%%%%%%%%%%%%%%%%%%%%%%%%%%%%%%%%%%%%%%%%%%%%%%%%%%%%
%Paragraph 2: Our research question, methodology, findings
%%%%%%%%%%%%%%%%%%%%%%%%%%%%%%%%%%%%%%%%%%%%%%%%%%%%%%%%%%%%%%%%%%%%%%
This paper uses the annual stocks of firearm licenses and the total number
of registered and restricted firearms collected by the Royal Canadian 
Mounted Police (RCMP) between 2013 and 2019 to investigate the relationships 
between legal guns variables and homicides, suicides, and crime.  This new data highlights 
the rapid growth in legal ownership in this period. Restricted firearms licenses grew 
by 72.88\% in this period while the number of registered handguns grew by 52.19\%.  
This growth occurred while the adult population of Canada grew by 8.3\%.\footnote{Authors'
calculations based on RCMP data and Statcan Table: 17-10-0005-01.}

All of our analysis is specific to the Canadian context, which has particular laws 
surrounding firearms and which do not necessarily exist elsewhere.  Section 
\ref{section:backg} provides more details, but to highlight some differences,
firearms licenses are mandatory for anyone to legally possess or acquire a firearm 
in Canada.  Further comparisons between the Canadian and American regulatory landscapes
can be found in Table \ref{tab:can_us_diff}. This licensing requirement can be understood
as the extensive margin of legal firearm ownership, as it provides individuals with the 
ability to purchase firearms. Registered and restricted firearms are a subset of all 
firearms which can be legally bought in Canada, as legally purchased unrestricted firearms
no longer need to be registered.

Although they do not include all legal firearms, we have data on the number of all 
publicly registered handguns and restricted rifles.  Taken together, these data align
closely with the firearms affected by the handgun freeze, while for the buyback they 
reflect only the portion of the affected long-gun stock captured in our restricted 
firearms measures. These variables improve upon previous studies, which often rely 
on proxy measures for firearms due to data limitations.

%%%%%%%%%%%%%%%%%%%%%%%%%%%%%%%%%%%%%%%%%%%%%%%%%%%%%%%%%%%%%%%%%%%%%%
%Paragraph 2: Methodology and Findings 
%%%%%%%%%%%%%%%%%%%%%%%%%%%%%%%%%%%%%%%%%%%%%%%%%%%%%%%%%%%%%%%%%%%%%%
We combine our licenses and guns variables with both publicly available 
data on deaths, homicides and firearms-related crimes and administrative 
data on deaths in Canada. Our main model specifies the prevalence of our
outcome variables (deaths, crimes, \textit{etc.}) as a function of the 
prevalence of our firearms variables.\footnote{In what follows, \textit{prevalence} 
indicates values are rates per $100,000$ unless otherwise noted.} We use a linear 
fixed effects model to control for unobserved heterogeneity across time and Census
Metropolitan Areas (CMAs) and Census Agglomerations (CAs).%
\footnote{A CMA can be thought of as a Metropolitan Statistical Area (MSA)%
; a CA can be thought of as a Core Based Statistical Area (CBSA).} %
CMAs and CAs are largely urban parts of Canada and so our analysis currently omits 
more rural parts of Canada.

%
% RDC Deaths Data 
%
We accessed Statistics Canada's \textit{Canadian Vital Statistics Death
(CVSD)} database which provides restricted data on individuals' leading 
cause of death. This allows us to aggregate deaths into those involving
firearms, as well as the intent of deaths involving firearms and which
type of firearm is thought to have caused the death. We further
aggregate these deaths to their CMA and CA at the annual level to match 
our firearms dataset. 

When modelling the impacts of the prevalence of registered firearms or 
licenses separately from one another with CMA and CA fixed effects, we fail 
to detect any effects on firearms-related deaths, firearms-related 
suicides, firearms-related assaults or any of those deaths where a 
specific firearm was suspected of being used in a death in Canada between 
2013 and 2019. Modelling the prevalence of licenses and restricted firearms 
in the same model still yields no effects across all of our dependent variables.

We also estimate a model which regresses our different firearms-related 
deaths with all of our guns variables. These models disaggregate both types of 
license holders into Possession and Acquisition Licenses (PALs) and Restricted 
Possession and Acquisition Licenses (RPALs). We additionally consider three 
categories of  restricted firearms: handguns, rifles and a residual category 
of `other  guns'. All firearms-related deaths are statistically significant and 
positively related to other guns, but not the other two categories. Both license types 
as well as handguns and rifles are statistically insignificant at traditional levels. 
The coefficient estimates suggest increasing other guns by about $50$ (per $100,000$) 
would increase the firearms-related deaths by $0.1$ (per $100,000$). When decomposing 
total firearms deaths into those  whose intents are assaults or self-harm, a similar 
story emerges. Handguns,  PALs and RPALs are unrelated to firearms deaths whose 
intent is assaults,  self-harm or unknown. Coefficient estimates suggest increasing the 
other guns variable by about $139$ ($83$) per $100,000$ would increase 
firearms-related assault (self-harm) deaths by $0.1$ (per $100,000$).
Rifles are marginally significant and negatively related to 
self-harm deaths, where coefficient estimates suggest increasing rifles 
by about $3,845$ (per $100,000$) would reduce self-harm related deaths by 
$0.1$ (per $100,000$). There seem to be no effects of firearms on deaths whose 
intent is unknown, across all of our models in Canada over this time period.

We use the same model but change our outcome variable to firearms-related deaths 
believed to be caused by a specific type of firearm (rifle, handgun, \textit{etc.}). 
While effects are statistically significant, 
the coefficient magnitudes are often too small to be of practical economic 
importance. The coefficient estimates suggest that an increase of $5,800$ handguns 
(per $100,000$) is associated with a $0.1$ increase in handgun-related deaths 
(per $100,000$). Given that the sample mean for handguns is approximately $3,200$ 
(per $100,000$), this effect appears quite substantial. In contrast, increasing the 
prevalence of restricted rifles by about $22,220$ (per $100,000$) is associated with 
a $0.1$ decrease in total handgun-related deaths (per $100,000$). Again, the sample mean 
of restricted rifles is approximately $400$ (per $100,000$). Rifles and other guns are 
statistically significant and negatively related to rifle deaths. Coefficient estimates 
suggest a $0.1$ (per $100,000$) decrease in deaths associated with approximately $4,350$ rifles 
and $190$ other guns (per $100,000$). Other guns and licenses are statistically significant 
and positively related to firearms deaths with unspecified firearms. 
Coefficient estimates suggest increasing unspecified firearms deaths by
$0.1$ (per $100,000$) from increasing other guns by about $40$ (per $100,000$). 
Increasing PALs by about $526$ (per $100,000$) would translate to increasing 
unspecified firearms deaths by about $0.1$ (per $100,000$). There
does not appear to be a relationship between uncategorised firearms deaths
and our guns variables across all of our models. 

%
% CRIMES DATA 
%
We also estimate the effects of our firearms variables on firearms
related crimes. We first standardize our firearms variables to more
easily interpret our coefficients. Using the (standardized) prevalence 
of all licenses and all registered and restricted firearms as independent 
variables, we see that licenses appear to be 
decreasing documentation issues (coefficient equalling $-1.56$, 
significant at the 5\% level) and also total homicides (coefficient 
equalling $-0.33$, significant at the 1\% level). Breaking out total 
homicides into first degree murder, second degree murder and manslaughter, 
we find positive effects on first degree murders (coefficient equalling 
$0.53$, significant at the 10\% level), negative effects on second
degree murders (coefficient equalling $-0.88$, significant at the 5\% 
level), and no effects on manslaughter. There are no effects 
of total restricted firearms on any of our crime variables. 

We then break out both our licenses into PALs and RPALs and our 
firearms variables into rifles, handguns, and other guns. The results
are heterogeneous across types of licenses and types of firearms. Rifles
appear to be increasing total firearms violations (through firearms 
discharge, improper use of firearms and improper storage offences).  ``Total 
Firearms Violations'' refers to UCR incidents involving firearm use, discharge, 
or pointing, while documentation and improper storage offences are treated as 
separate categories. Handguns have null effects on total firearms violations 
(decreasing using firearms and improper storage offences but increasing documentation
issues). 

Total homicides are unrelated to PALs, RPALs, handguns, rifles, and
other guns. However, there is heterogeneity amongst the different types of
homicides and our guns variables. Handguns and PALs are positively related to first degree
murders; rifles and PALs are negatively related to second-degree
murders. Rifles, other guns and PALs are positively related to 
manslaughters but RPALs are negatively related to manslaughters. 

%%%%%%%%%%%%%%%%%%%%%%%%%%%%%%%%%%%%%%%%%%%%%%%%%%%%%%%%%%%%%%%%%%%%%%
% 
One possible explanation for our muted, or null, results is that variation 
in legal gun owners need not affect firearms-related crime and homicides 
the same way as illicit gun use, as discussed in Appendix Figure \ref{fig:DAG}. 
If illicit gun use has a stronger 
association with firearms-related crime and homicides, legal guns may 
not be the margin policymakers ought to be concerned 
with. If illicit firearms are what 
reduces public safety, handgun bans and or gun buyback 
programs need not be associated with improved public safety. In the US, \citet{Khalil2017} 
finds increasing crimes are associated with lagged values of illegal 
firearms flows (their proxy being the number of reported stolen 
firearms in policy jurisdictions) but not associated with lagged values 
of legal firearms flows (their proxy being percentage of gun suicides). 

Currently, we do not have a similar measure for illegal firearms and 
their impacts on crime since data tracking reported stolen or missing 
firearms is not readily available in Canada. More importantly, evidence suggests that 
a large share of illegal firearms in Canada originate from smuggling from the United 
States, particularly in large urban centres.\footnote{For example, nearly 90\% of 
crime guns seized in Toronto last year were traced back to the U.S. \citep{petz2025guns}.} 
However, historical measures suggest quantities of illegal firearms to be small relative 
to the legal stock, though they do not provide a direct measure of the total illicit firearm stock.
During the 2019 fiscal year the Canadian Border Services Agency seized 
753 firearms.\footnote{See the statistics on 
\url{https://www.cbsa-asfc.gc.ca/security-securite/seizure-saisie-eng.html}} 
While this value is small relative to the $1,159,477$ registered and restricted firearms
in our data as of December 2019, it represents a limited indicator of illicit activity 
rather than a direct measure of the overall illegal firearm stock. As another example, the ratio 
of total firearms reported missing or stolen to the cumulative total 
of unique firearms registered in 2004 was $0.000806$ (=$5,631/6,984,485$) 
\citep{Hung2006}.\footnote{The total number of unique firearms is 
substantially different than the number in our data set due to institutional changes regarding which 
firearms must be registered over time.} Moreover, the object of this 
paper is to estimate the impacts of legal gun ownership on firearms-related
deaths since these are the policies being considered.

%%%%%%%%%%%%%%%%%%%%%%%%%%%%%%%%%%%%%%%%%%%%%%%%%%%%%%%%%%%%%%%%%%%%%%
Insufficient data on direct measures of legal firearms, especially in the US,  leads
many researchers to use proxy variables that are likely correlated with the presence
of firearms and their owners. Examples of proxies include the sales of gun enthusiast
magazines and National Rifle Association (NRA) gun shows \citep{Duggan2001}, the 
percentage of gun suicides \citep{Cook2006, Khalil2017} or the number of background
checks \citep{Lang2013}. While more direct measures of legal firearms supply like the 
density of gun dealerships \citep{johnson2024gun} and firearms sales in Pennsylvania
\citep{Johnson2022} are beginning to enter the literature, data on gun ownership and 
total guns data for any region are still hard to come by. Existence of such data 
represents an improvement on this literature in general. Additionally, since our data
on license holders and numbers of legally owned firearms is administrative, concerns 
over attenuation bias ought to be reduced. 

Much of the previous research uses (proxy) flow measures of firearms and firearm owners.
Our work is able to measure both the annual change (flow) and the stock since we have the
annual total number of licenses and registered and restricted firearms in the legal 
market.\footnote{Unregistered firearms which are classified as restricted are illegal 
firearms, by definition.} While we only observe restricted firearms (a subset of the 
total legal firearms in Canada), we perfectly measure the number of issued licenses. 
Moreover, since all license owners are subject to a background check when applying for
a license, and continuously while licenses are valid, we have the stock of all individuals
who currently pass background checks.

%%%%%%%%%%%%%%%%%%%%%%%%%%%%%%%%%%%%%%%%%%%%%%%%%%%%%%%%%%%%%%%%%%%%%%
% other measures of firearms
More recent studies use plausibly exogenous variation to understand important relationships
with firearms and to measure the efficacy of various policy changes. \citet{colmer2026access} 
argue that daily temperatures cause higher incidences of homicides and assaults with larger 
effects for states with less restrictive right-to-carry laws. \citet{cook2024dona} find a 
similar relationship using gunshots identified by audio monitors in Chicago.  \citet{Carr2018} 
argue that youth curfew laws aimed at reducing crimes, controlling for fired shots, simply shift
when crime takes place. This type of variation likely affects firearm owners and the environments
they face, and is often used to indirectly measure firearm use. The nature of our data, being
annual and aggregated to a much larger geographic region, does not permit us to use a similar 
research design.

%%%%%%%%%%%%%%%%%%%%%%%%%%%%%%%%%%%%%%%%%%%%%%%%%%%%%%%%%%%%%%%%%%%%%%
%LIT REVIEW: CANADA 
We additionally contribute directly to the literature focusing on firearms-related harms in Canada. 
Evidence-based research on firearms in Canada is comparatively small and focuses predominantly 
on gun-related violence, like homicides and suicides which is due in part to a lack of relevant data.%
\footnote{For example, \citet{Fergunson2019} finds only thirteen peer-reviewed publications on 
firearms-related research in Canada between 2000 and 2019.} %
\citet{Leenaars_1994} is an early analysis of the impact of gun control on homicides in 
Canada while \citet{Dandurand1998} summarizes firearms use and gun-related violence prior
to much of the legislative changes which have come before the mid-90s. \citet{Gagne_2010} uses more
restrictive firearms legislation in Quebec during 1991 to show a decrease in male suicides involving
a firearm. Reviewing the Canadian evidence on suicide, \citet{langmann2021suicide} argues that 
although some firearms legislation is associated with reductions in firearm suicides, the evidence 
does not demonstrate a corresponding reduction in overall suicide rates. There has been increasing 
interest in firearms-related research following  several recent Canadian federal election campaigns
promising improved public safety with regard to firearms. Using a difference-in-differences analysis
of Canadian mortality data from 1981 to 2016, \citet{Langmann2020} finds no evidence that firearms
legislation reduced overall suicide or homicide rates and no association between provincial 
licensing rates, used as a proxy for firearm prevalence, and suicide rates. However, unlike our 
study, it does not use direct data on the stock of firearms. More recently, 
\citet{langmann2025critical} challenges claims that firearms legislation reduced firearm mortality
in Canada and Australia, arguing that the existing evidence is sensitive to methodological choices
and does not establish a clear causal relationship. Finally, research by \citet{Schwartz2021} 
demonstrates how an emerging social identity through gun ownership is associated with increased
political activity.

%%%%%%%%%%%%%%%%%%%%%%%%%%%%%%%%%%%%%%%%%%%%%%%%%%%%%%%%%%%%%%%%%%%%%%
% Lit Review: Contribution to Literature. UNITED STATES, General. 
This paper broadly contributes to the literature by asking how guns impact public safety. While 
research examining the relationship between firearms and gun-related deaths exists internationally, 
research tends to focus on the United States.%
\footnote{\citet{Killias_2012} provides a European overview of the association between firearm 
availability and murders.} %
\citet{Moody_2010} provides an overview of the relationship between handguns and homicide in the 
US. \citet{Brent_2001} offers an early review of the relationship between firearm ownership and 
suicides, finding the presence of a firearm in the home increases the risk of suicide. 
\citet{Miller_2015} examine the relationship between firearm ownership and suicides at the city 
level in America.  Using survey data on ownership, they find that areas with higher firearm 
ownerships have higher rates of suicide using a firearm, but no impact on suicides not using a 
firearm. \citet{Anestis_2018} find a similar relationship at the state level. \citet{Knopov2019}
find that states with higher proportions of gun ownership in households are associated with higher
youth suicide rates.

%%%%%%%%%%%%%%%%%%%%%%%%%%%%%%%%%%%%%%%%%%%%%%%%%%%%%%%%%%%%%%%%%%%%%%
%Paragraph 5: roadmap
%%%%%%%%%%%%%%%%%%%%%%%%%%%%%%%%%%%%%%%%%%%%%%%%%%%%%%%%%%%%%%%%%%%%%%
This paper is structured as follows. Section \ref{section:backg} gives a background on
gun ownership in Canada. Section \ref{section:concept} describes the environment and 
channels through which firearms policy can act. 
Section \ref{section:data} describes our dataset before discussing our model and 
estimation strategy in Section \ref{section:strategy}. Section \ref{section:results} 
presents our results and Section \ref{section:concl} concludes.

%Background
\section{Background: Licensing and Recent Legislation} \label{section:backg}

%%%%%%%%%%%%%%%%%%%%%%%%%%%%%%%%%%%%%%%%%%%%%%%%%%%%%%%%%%%%%%%%%%%%%%
%types of firearms
Regulated firearms in Canada fall into three distinct categories: non-restricted, 
restricted, and prohibited.\footnote{A detailed discussion of the differences can be
found here \url{https://www.rcmp-grc.gc.ca/en/firearms/classes-firearms}.} %
Generally speaking, prohibited firearms are fully-automatic firearms, 
handguns with short barrels, small calibre handguns, and firearms deemed 
prohibited.  Restricted firearms are non-prohibited handguns and rifles which
are deemed to be restricted. Non-restricted firearms are firearms which are neither
restricted nor prohibited. In effect, non-restricted firearms are mostly shotguns and 
rifles.  Generally speaking, the ability to own or purchase different types
of firearms is largely dictated by Canadians' licenses. Individuals without a 
valid firearms licence cannot legally possess or acquire firearms.

\subsection{Gun Licensing in Canada: PALs and RPALs}

%%%%%%%%%%%%%%%%%%%%%%%%%%%%%%%%%%%%%%%%%%%%%%%%%%%%%%%%%%%%%%%%%%%%%%	
% Licensing in Canada 
%%%%%%%%%%%%%%%%%%%%%%%%%%%%%%%%%%%%%%%%%%%%%%%%%%%%%%%%%%%%%%%%%%%%%%
The Canadian firearm regulatory landscape is complex. Legislation (as of
2022) largely limits the ownership of firearms, carrying firearms 
in public, using firearms in the commission of crimes, and buying, 
selling, importing or exporting firearms. The Canadian  
firearm licensing system is an involved system of training, testing, 
and constant background-checking which grants and tracks individuals'
licenses. Licenses can be regarded as exceptions to the overall 
prohibition of firearms ownership in Canada.

%%%%%%%%%%%%%%%%%%%%%%%%%%%%%%%%%%%%%%%%%%%%%%%%%%%%%%%%%%%%%%%%%%%%%%
% Licensing 
There are two general types of firearm licenses in Canada: Possession and
Acquisition licenses (PALs) and Restricted Possession and Acquisition 
Licenses (RPALs). These licenses differ in the types of firearms that 
they allow an individual to acquire. %
PALs allow for the purchase and acquisition of non-restricted firearms. An RPAL 
can be understood as an extension of a PAL, as it confers the privileges of a 
PAL and allows purchasing or acquiring of restricted firearms. The general 
public is not able to obtain a license that allows for ownership of prohibited 
firearms.

%%%%%%%%%%%%%%%%%%%%%%%%%%%%%%%%%%%%%%%%%%%%%%%%%%%%%%%%%%%%%%%%%%%%%%
%
There are many steps to obtain and maintain a firearms license.  Individuals
can obtain  a PAL by passing the Canadian Firearms Safety Course (CFSC), which 
includes in-class instruction, a practical test and a written test. Applicants who
pass the CFSC are then vetted through the Royal Canadian Mounted 
Police (RCMP) using a background check and declaration of 
personal information and personal history.%
\footnote{Appendix Figure \ref{apndx_infoSheet} shows the opening page of the
Information Sheet, while the Personal History section is shown in Appendix 
Figure \ref{apndx_personalHistory}.} %
The Personal History section requires applicants to answer questions 
about their previous criminal history, mental health concerns, separation 
from a partner in the previous two years, or if they've been reported 
for violent behaviour. Having any of these concerns does not preclude 
having a PAL but may require additional investigation for the application
to be approved. Finally, two references (neither of which can be an applicant's 
partner) who have known the applicant for at least three years must sign that they
``know of no reason why, in the interest of safety of the applicant 
or any other person, the applicant should not be given a licence to possess 
and acquire firearms.''

%%%%%%%%%%%%%%%%%%%%%%%%%%%%%%%%%%%%%%%%%%%%%%%%%%%%%%%%%%%%%%%%%%%%%%
%
RPALs can be obtained only after successfully passing the Canadian 
\textit{Restricted} Firearms Safety Course in addition to the CFSC. 
RPAL applicants attend in-class sessions, take practical and written 
tests, and apply for their upgraded licence through the RCMP in a process
akin to the PAL.

%%%%%%%%%%%%%%%%%%%%%%%%%%%%%%%%%%%%%%%%%%%%%%%%%%%%%%%%%%%%%%%%%%%%%%

Total costs of each course (in-class portion and RCMP application) are
less than $\$200$ (Nominal, 2022, Canadian Dollars). The in-class portion
is often completed in a weekend, with the processing time for applications
ranging from two to six months.\footnote{An example of a firearms course can
be found here: %
\url{https://ottawafirearmsafety.ca/pal-application/}.%
} %
Licences must be renewed once every five years at a cost of about $\$100$ 
(Nominal, 2022, Canadian Dollars).

%%%%%%%%%%%%%%%%%%%%%%%%%%%%%%%%%%%%%%%%%%%%%%%%%%%%%%%%%%%%%%%%%%%%%%
%
In addition to the licensing requirements, the Chief Firearms Officer (CFO) of each 
province has discretion over whether to approve the transfer of a firearm after a 
sale.  In every province except for Ontario, CFOs also require RPAL holders to 
have a current membership to an accredited firing range in order to acquire
a restricted firearm.%
\footnote{%
See \url{https://www.rcmp-grc.gc.ca/en/firearms/classes-firearms} %
for additional information.%
}
Given that transfers need to be approved by a CFO for every acquisition of a restricted 
firearm, we have very detailed information about the geographic dispersion of 
firearms. Similarly, license holders must provide their address when they first 
apply and update this immediately as they move. Thus, there is also up-to-date and
detailed information about the spatial distribution of license holders.  Permission is
also needed to relocate restricted firearms from the old residence to the new one. 

\subsection{Recent History of Firearms Legislation in Canada}

%%%%%%%%%%%%%%%%%%%%%%%%%%%%%%%%%%%%%%%%%%%%%%%%%%%%%%%%%%%%%%%%%%%%
%
Canadians have been subject to major legislative changes regarding 
gun ownership since the early 1990s. 
Bill C-17 and the \textit{Firearms Act} (Bill 
C-68) were passed in 1991 and 1995, respectively, each introducing
many aspects of the modern licensing program in Canada \citep{RCMP2020, 
	CanadaEncyc2021}. The former introduced a more stringent vetting 
process for licensing such as spousal endorsement and safety training, 
new and amended legal penalties such as safe storage laws and minimum 
sentences, and prohibited certain types of guns \citep{Langmann2012}. 
Bill C-17 was rolled out through the 1990s \citep{RCMP2020} while 
the \textit{Firearms Act} was being conceived and would ultimately 
alter aspects of gun legislation in Canada. Although 
coming into effect in 1998, the former Firearms and Acquisition 
Certification (FAC) system would be replaced by the Possession and 
Acquisition Licence system on January 1st, 2001  \citep{RCMP2020}. The PAL 
licensing requirement came into force on January 1, 2001, while 2003 marked
the deadline for registering all firearms, particularly non-restricted long guns.

%%%%%%%%%%%%%%%%%%%%%%%%%%%%%%%%%%%%%%%%%%%%%%%%%%%%%%%%%%%%%%%%%%%%%%
%
Most of the major changes from firearms legislation in the 1990s 
continued to come into effect in the early 2000s. Considerable debate
regarding non-restricted firearms registration led to 
legislative proposals aimed at removing the registration. This would 
be proposed as Bills C-21, C-24, and C-19 in 2006, 2007, and 2011 
respectively \citep{RCMP2020}. Where the first two failed, the third
succeeded, and came into effect on April 5th, 2012. Bill C-19 in 2012 
``... eliminate[d] the registration of non-restricted firearms and 
erase[d] the data from the registry'' \citep{CanadaEncyc2021}.%
\footnote{Quebec challenged this in the Supreme Court before 
	ultimately losing on March 27th, 2015, after which their records
    were also destroyed. 
}
The records were thus destroyed by October 2012. Importantly for us, 
detailed data on the registry of guns are available only for years 
following 2012.

%%%%%%%%%%%%%%%%%%%%%%%%%%%%%%%%%%%%%%%%%%%%%%%%%%%%%%%%%%%%%%%%%%%%%%
%
The \textit{Common Sense Firearms Licensing Act}, Bill C-42, was 
enacted on July 31st, 2015 and came fully into force September 2nd of 
the same year \citep{Canada2015_A, Canada2015_B}. Bill C-42
transitioned all previously grandfathered Possession Only License holders 
to a PAL license, made in-class training for first-time licence 
applicants mandatory, and allowed restricted firearms to be 
transported without seeking authorization for pre-approved locations 
like firing ranges, gun repair stores, and to an international border, 
amongst other changes. Our data on licences show a one-time increase between 2015 
and 2016 of PALs which we believe are the POLs being grandfathered in. A 
possession-only licence allowed for ownership of existing firearms, but not the
ability to acquire new ones. Ironically, the handgun freeze can be seen as 
downgrading an RPAL to a POL for existing owners of handguns.

%%%%%%%%%%%%%%%%%%%%%%%%%%%%%%%%%%%%%%%%%%%%%%%%%%%%%%%%%%%%%%%%%%%%%%
% 
Bill C-71 passed on June 21st 2019, furthering precautionary measures like lifetime 
background checks (up from only the previous five years) and 
making authorization for transportation required (except for 
trips to a firing range) \citep{Canada2019, RCMP2021}. As with the 
previous legislation, there was a delay between when passed and when 
implemented, with both background checks and transportation changes 
starting in 2021.

%%%%%%%%%%%%%%%%%%%%%%%%%%%%%%%%%%%%%%%%%%%%%%%%%%%%%%%%%%%%%%%%%%%%%%
% Nova Scotia shooting, rifle ban, handgun ban
In April 2020, a gunman in Nova Scotia went on a mass killing leading to the
death of 22 individuals and the perpetrator. This event, among others, led
to the prohibition of 1500 semi-automatic rifle types by the government. Many
of these firearms were previously only legally accessible \textit{via} PAL or RPAL 
 holders.\footnote{An amendment to expand the list of prohibited semi-automatic
assault rifles was proposed in late 2022. The new list included popular types of 
hunting rifles. Several groups, including first nations groups, protested and 
the amendment was withdrawn.} The federal government further bolstered their stance on
restricting access to firearms in 2021 by proposing legislation that 
would allow municipalities to ban handguns. By September 2022, the 
semi-automatic assault-style rifle buyback program had classified 
weapons that it intended to buy back. In October 2022, the Federal 
government froze the market on handguns, no longer allowing handguns
to be bought, sold or transferred, except for exempt individuals and
businesses.

%%%%%%%%%%%%%%%%%%%%%%%%%%%%%%%%%%%%%%%%%%%%%%%%%%%%%%%%%%%%%%%%%%%%%%
%Potential Outcomes of The recent policy changes 
Handgun freezes in Canada apply at the extensive margin to a unique set of individuals. 
Specifically, handgun freezes impact individuals who 
have an RPAL (restricted license) so can own a handgun 
but are yet to purchase a handgun. The handgun freeze is thus quite narrow in scope.
People who currently possess legally owned handguns are no longer allowed to 
sell or transfer any handguns, but may still possess them. While the recent bill does not 
limit individuals from applying for RPALs, the benefits of an RPAL over a PAL
are significantly diminished. The handgun freeze does not directly target unlicensed 
(illicit) gun owners since all non-licensed individuals are already 
prohibited from owning a handgun.  Similarly, the buyback affects current 
legal owners of firearms that are now prohibited but were previously 
non-restricted, restricted, or in some cases not classified as firearms. 
The buyback does not apply to people who illegally own these firearms.  Recall,
the freeze has become law but the prohibited firearms buyback has not yet been 
implemented \citep{psc2024commonsense, psc2025compensation}.

\subsection{Important Differences between Canada and the US \label{section:US_Can_Diff}}
%
%%%%%%%%%%%%%%%%%%%%%%%%%%%%%%%%%%%%%%%%%%%%%%%%%%%%%%%%%%%%%%%%%%%%%%
%Paragraph: Comparison to the US? 
Much of the past literature on the relationship between gun ownership 
and homicides has used American data. There
are reasons to suspect that past results are not easily transferable
to Canada. Table \ref{tab:can_us_diff} notes the differences between
regulating firearms in Canada and the US as of May 2022. The only 
commonality is that both countries require individuals who own a 
firearm to be older than 18. In Canada, background checks, training 
and licensing are mandatory for ownership of firearms and enforced 
at the national level. In contrast, only some states in the US make
these background checks, training and licensing mandatory. In Canada, 
the following regulations are enforced nationally while comparable regulations 
are not implemented uniformly nationally in the US: (1) firearms registration; 
(2) restricted ownership of handguns; (3) prohibition of most 
magazine-fed semi-automatic rifles; (4) prohibition of short-barrelled
handguns. Instead, firearm regulations in the United States vary across states
and local jurisdictions, resulting in a patchwork of policies with differing levels
of stringency. Additionally, Canadians are generally not able to
open- or concealed-carry handguns whereas some American states permit either open-
or concealed-carry of handguns.%
\footnote{%
Technically, an RPAL holder can apply for an Authorization to Carry 
(ATC) for personal protection reasons. However, according to one
\textit{Access to Information and Privacy} %
request in 2018, only two individuals in Canada had a valid ATC. See %
\url{https://thegunblog.ca/2018/11/08/two-canadians-have-authorization-to-carry-guns-filing-shows/}.
} %
All Canadians are also required to comply with safe storage laws, including
the use of trigger locks, and must ensure that firearms are unloaded when stored. 

In Canada, there is no Second 
Amendment (``Right to Bear Arms'') as there is in the 
United States. Firearms in Canada are federally regulated, allowing 
much swifter changes in the legal landscape with regards to firearms. 
This also impacts our analysis, since there is no heterogeneity 
at a sub-national level due to legal changes.\footnote{There was previously a 
federal election campaign pledge in 2019 to allow municipalities 
(census subdivisions, which can be thought of  as counties) to ban handguns.}  

Many gun control advocates in the United States have called for some or
all of these measures.\footnote{See for example the list of `solutions' from the 
advocacy group \textit{Everytown} \url{https://www.everytown.org/solutions/}.}  
Thus the current regulatory environment of Canada provides an interesting case 
study for America. Some states and cities have policies that partially resemble 
aspects of Canadian regulations, but again there is no comparable 
nationwide framework.

As previously discussed, there is generally a lack of precise
data on firearms ownership in the United States, so many studies have 
used proxies for ownership.  Consequently, most studies involve measures
that include both legal and illegal firearms. In this study we are able 
to precisely measure legal (registered and restricted) firearms at the 
local level. 

\section{Conceptual Framework} \label{section:concept}

%%%%%%%%%%%%%	
%Paragraph 1: Describing the Figure  
%%%%%%%%%%%%%
Appendix Diagram \ref{dia:arms_markets} helps explain our conceptual framework 
about the relationship between legal and illegal markets for firearms and various
uses. \textit{Nature} randomly assigns individual $i$ their proclivity for firearms
ownership ($\theta$). Each individual selects into one of four broad categories 
given an exogenous regulatory environment ($R$). The regulatory environment is taken
as given in the individual decision problem, while policies such as background checks,
training, and penalties capture comparative statics across regulatory regimes. 
Individuals may not want to access firearms at all which we call the ``Never'' type
or $\theta_{0}$. Otherwise, those wishing to access firearms may only be willing to
do so through legal channels (``Legal'' type or $\theta_{L}$); some may be only 
willing (or able) to access through illicit channels (``Illicit'' type or 
$\theta_{I}$); and some may be willing to access firearms either legally or 
illegally (``Either'' type or $\theta_{E}$). Thus, two markets form: the formal 
market ($M_{F}$) for firearms which is regulated and an unregulated, informal 
market ($M_{\neg F}$), which is subject to legal penalties. The formal market is 
accessed only by the ``Legal'' and ``Either'' types, while the informal market can 
be accessed by ``Illicit'' and ``Either'' types. The dashed lines starting from 
the ``Either'' type and entering into both markets signify that the individual can
be swayed into either market. Once individual $i$ accesses firearms through their
respective market, they can choose how to use firearms. In our model, a person's 
type determines their motivation for use. ``Type'' reflects underlying traits and 
constraints (such as intended use and willingness to engage with legal institutions) 
that jointly shape both market access and firearm use. There are three firearms 
uses: for sport, for self-harm, and for crime.\footnote{The diagram could include
self-defence. In Canada, firearms ownership for purposes of self-defence is not 
generally permitted through legal channels.}  While either market may yield any 
of the uses of firearms, we show two dashed lines (one from the informal market 
to sport; the other from the formal market to crime) which we believe to be possible
but unlikely. Again, this is due to one's motivation for use.

%%%%%%%%%%%%%	
%Paragraph 2: Describing the Policy Channels 
%%%%%%%%%%%%%

For many of these linkages there exist potential incentive schemes or policies capable 
of influencing the regulatory environment ($R$). For example, those wishing to access 
through legal channels may be subject to background checks, mandatory training and 
licensing. That is, the edges entering the formal market can be influenced by the 
stringency of regulations directed at access. After accessing firearms through the 
legal market, firearms owners may have their use restricted to sport. In so doing, 
additional regulations such as safe storage laws or the ability to monitor where 
firearms are transferred would result in non-increasing self-harm and crimes. The 
presence of monitoring in the formal market may help reduce self-harm by enabling 
oversight that can restrict firearm ownership when necessary.

For those owning firearms in the informal market, there may be substantial legal 
penalties associated with possessing firearms illegally. These penalties influence 
the edges entering the informal market or the edges entering any of the downstream 
uses. That is, there may be direct penalties, like illegally possessing a firearm or 
there may be deterrents for using firearms in crimes which affect the extensive 
margin decision to possess firearms in the first place. These penalties are 
likely to compound if one was using firearms in accompanying crimes like 
robbery, assault or homicide. 

%%%%%%%%%%%%%	
%Paragraph 2: Relevance to Context 
%%%%%%%%%%%%%

Policymakers are able to alter the environment described in Appendix Diagram
\ref{dia:arms_markets} in multiple dimensions. Policymakers could reduce the 
incentives of owning firearms in the informal sector by increasing penalties 
for getting caught. This would restrict the number of illicit individuals and
may deter ``either''-typed individuals into the legal market or choosing not 
to own. Alternatively, policymakers may increase the monitoring of the formal 
market, decreasing the likelihood for self-harm and crime. Monitoring costs
``legal'' or ``either'' types; they may stop purchasing altogether or
substitute towards the informal market. 

The framework clarifies the avenues through which policymakers can influence 
the firearms landscape. If policymakers would like to reduce the informal market 
and crime, they can do so directly. Alternatively, if policymakers want to reduce 
self-harm associated with formal market firearms owners, they can do so there. 
This framework also suggests why our analysis yields many null results: individuals
wanting firearms for criminal purposes are unlikely to enter the formal market.

% Data

\section{Data}\label{section:data}
\subsubsection*{Annual Gun Licenses and Types of Guns}

We acquired data on gun ownership in Canada \textit{via} request 
under the \textit{Access to Information and Privacy (ATIP) Act}. 
The Royal Canadian Mounted Police (RCMP) provided data characterizing 
two major aspects of gun ownership in Canada: the number of firearms
recorded in the federal registry (primarily restricted and prohibited firearms) 
and the number of valid gun licenses.%
\footnote{We also requested one dataset about the sellers of guns 
-- counts of licensed \textit{dealers} grouped by postal code -- in 
Canada. Since this data is a cross-section for 2020, it can be used 
only for descriptive statistics and figures. Moreover, gun dealers 
are allowed to sell firearms to license holders across the country 
\textit{via} mail. In this way, the availability of local dealers 
is not as meaningful as in places like the United States \citep{johnson2024gun}.
}
Throughout this text, any explicit mention of the counts of firearms,
like handguns or rifles, should be understood to refer to the total
\textit{registered and restricted guns}. Importantly, this does not 
measure unrestricted firearms, like most shotguns or rifles, and any 
illicit firearms or unlicensed owners. That is, we have administrative 
data on the formal market and no data from the informal one.

Counts of guns are separated into the following categories:
handguns, rifles, shotguns, submachine guns, machine guns, air
pistols, combination guns and commercial versions of guns. Handguns
and rifles make up the majority of these firearms. 

We aggregate shotguns, submachine guns, machine guns, air pistols, 
combination guns and commercial guns into a category we call ``other guns.'' 
Because this category comprises a relatively small share of total registered 
firearms and includes several types that are rare or inconsistently reported, 
we place less emphasis on results derived from this group in our analysis. We
note that the empirical ``other guns'' category is a residual category that 
includes some but not all firearms classified as prohibited under Canadian 
law, and therefore does not correspond one-to-one with the legal ``prohibited''
firearm class.

Counts of licensed gun owners fall into one of two major categories: 
those with a valid possession and acquisition firearm license (PAL) 
and those with valid \textit{restricted} possession and acquisition 
firearm license (RPAL). Individuals are counted once in the category of their 
highest licence, so PAL and RPAL counts are mutually exclusive and 
sum to total licence holders.
Both gun owners and types of guns were given at the census
subdivision (CSD) level (which can be thought of as municipalities) 
and span the years between 2013 and 2019, inclusively. The data provided 
to us are as of December 31\textsuperscript{st} of each year. These 
municipalities likely follow the 2011 version of the Standard 
Geographic Classification (SGC).%
\footnote{% 
The ``likely'' caveat is a result of our trial-and-error merging
between various data sources. It was not made explicit to us 
the correspondence between the RCMP data and the SGC version. We do our
best to crosswalk the data we were given with much of our SGC 2016 data. 
In most cases, issues are at too fine a level to make a difference 
when aggregating counts to the appropriate geographic level (\textit{i.e.} 
provincial, census agglomeration, or census metropolitan area). 
See our Data Appendix \ref{sec:apndx_data} for an extensive
discussion.
} %
Not all of the 3753 municipalities are accounted for in our dataset, likely
from the amount of census subdivisions without any counts of guns 
or licenses. This seems to be worst in Quebec, where data for Montreal and the
surrounding municipalities as well as Quebec City are absent from the data 
given to us. In cases where there is no data on the census 
subdivision, we assume a value of zero for either licenses or 
registered and restricted firearms. %
In most of our analysis, we aggregate the licenses and firearms 
data to a more aggregate geographical area to match the other 
available data.

%%%%%%%%%%%%%%%%%%%%%%%%%%%%%%%%%%%%%%%%%%%%%%%%%%%%%%%%%%%%%%%%%%%%
\subsubsection*{Crosswalk between Census Subdivisions and Census 
	Metropolitan Areas, Census Agglomerations and Provinces}
	
%%%%%%%%%%%%%%%%%%%%%%%%%%%%%%%%%%%%%%%%%%%%%%%%%%%%%%%%%%%%%%%%%%%%
% Paragraph 1: What is it? 
Publicly available aggregate data in Canada is often reported at 
geographical levels called census metropolitan areas (CMAs) and less
frequently at census agglomerations (CAs).%
\footnote{Census metropolitan areas have at least 100,000 people living %
in the total area with at least 50,000 living in their ``core''. %
Census agglomerations have at least 10,000 in their core. Examples of % 
CMAs include Toronto, Montr{\'e}al, and Vancouver. Examples of CAs %
include: Sarnia, Cornwall, and Sault. Ste. Marie. %
} %
CMAs and CAs are much larger geographic regions
than a census subdivision. While census subdivisions do not map 
perfectly into CMAs and CAs, they do map reasonably well.%
\footnote{ This is mainly because they are two different
ways of splitting up Canada into smaller geographic units. We describe 
the process in greater detail in our data appendix \ref{sec:apndx_data}.
} %
We use a crosswalk between 
Provinces, CMAs, CAs, and census subdivisions to leverage our counts 
of guns and licensed owners with meaningful outcome variables like
deaths, homicides, crimes and 
covariates like population, employment, median income levels, and 
common benefits like the Canadian unemployment insurance 
(known in Canada as \textit{employment} insurance, or EI) and 
federal child benefits. After matching the guns data to the crosswalk, 
we aggregate our guns and licenses data up to the CMA/CA level. As a caveat, 
there are some CSDs which do not fit into any CMA or CA. The sample we are 
looking at largely contains urban Canada.

%%%%%%%%%%%%%%%%%%%%%%%%%%%%%%%%%%%%%%%%%%%%%%%%%%%%%%%%%%%%%%%%%%%%
\subsubsection*{Canadian Vital Statistics Death Database (CVSD)}
We accessed the Canadian Vital Statistics 
Death Database (CVSD), administrative deaths data through 
Statistics Canada's \textit{Research Data Center}.\footnote{All of our 
statistical analysis is performed in \textit{Stata}.} The dataset includes 
information on individuals, including the leading cause of death for an 
individual as well as an approximate location where the individual was 
residing at the time of death. The CVSD designates which ICD code best 
corresponds to their death, allowing us to aggregate individuals into types 
of deaths. The types of deaths we currently consider are: 
total firearms-related deaths; firearms deaths whose intent was 
designated as assaults, self-harm (suicides), or unknown; 
firearms deaths where the type of firearm used was a handgun, 
rifle, uncategorised or unknown. Firearm-related deaths are identified using 
the Classification of Death variables provided in the Canadian Vital Statistics 
Death Database (CVSD), which are constructed by Statistics Canada using ICD-10 
external cause codes. Intent categories (assault, self-harm, and unknown) and 
firearm type classifications (handgun, rifle/shotgun, and other or unspecified 
firearms) follow the standard Statistics Canada coding framework.

Individuals were aggregated to the year and CMA/CA level in which they died to 
better match our publicly available time-varying control variables. To 
aggregate to the CMA/CA, we used the 2016 \textit{Postal Code Conversion File} 
(PCCF), designating each individual to be living in the area given
by their single link indicator (SLI). This was necessary as the
CVSD does not have a uniform definition of census geographies 
across time. In total, we recover 1015 observations (CMA/CA-by-year) 
of the 1089 used with the publicly available data, with missing 
observations coming from CMAs or CAs which were promoted or 
demoted in 2016 due to changes in their population counts.

%%%%%%%%%%%%%%%%%%%%%%%%%%%%%%%%%%%%%%%%%%%%%%%%%%%%%%%%%%%%%%%%%%%%
\subsubsection*{Annual Deaths in Canada}

Data on the total number of deaths in Canada is gathered from two
sources. The first characterizes the leading cause of death at the
provincial level from \citet{StatCan_CauseDeath}. We use four large groups
which we label as suicides, homicides, accidents, and other causes 
of death.%
\footnote{%
Suicides are described as ``intentional self-harm (suicides)'' 
and include International Classification of Diseases (ICD) 10 codes 
[X60-X84, Y87.0]. Homicides are described 
as ``Assault (homicide)'' and include ICD-10 codes [X85 - Y09, Y87.1]. 
Accidents (unintentional injuries) include ICD-10 codes 
[V01-X59, Y85-Y86]. Other causes of death are a residual group. 
}

%%%%%%%%%%%%%%%%%%%%%%%%%%%%%%%%%%%%%%%%%%%%%%%%%%%%%%%%%%%%%%%%%%%%
\subsubsection*{Annual Counts of Incident-Based Crime Statistics}

Annual counts of various crimes come from \citet{StatCan_Crime} which
is an aggregate of the Uniform Crime Reporting (UCR) Survey. The data
include variables related to any violations, homicides, weapons 
violations, firearms use, and weapons trafficking, to name a few. 
These variables let us inspect if the distribution of guns across
Canada has any correlation with related offences. To the best of our 
knowledge, crime data for Canada during these years are only publicly 
available at the CMA level.

%%%%%%%%%%%%%%%%%%%%%%%%%%%%%%%%%%%%%%%%%%%%%%%%%%%%%%%%%%%%%%%%%%%%
\subsubsection*{Aggregate Taxfiler Data}

Taxfiler annual aggregate data at the CMA and CA levels come from 
\citet{StatCan_TaxFilers} and includes useful variables such as:
the number of people, the average age, the percent married (excluding 
common-law), average and median incomes, number of people who are 
using unemployment insurance, and number of people reporting federal
childcare benefits. These variables act as controls. 

%%%%%%%%%%%%%%%%%%%%%%%%%%%%%%%%%%%%%%%%%%%%%%%%%%%%%%%%%%%%%%%%%%%%
\subsubsection*{Descriptive Statistics CMA only and CMA/CA Datasets}

Tables \ref{tab:sstats_cmaca} and \ref{tab:sstats_cma} show the 
summary statistics of the variables which are used in our models 
involving deaths, crimes, guns variables, and additional covariates 
at the CMA/CA and only CMA levels, respectively. Our control variables 
show a reasonable variation associated with the continuum of city 
sizes across Canada: nearly half are female, the average age is 
about 41, and about 38\% of people are married. The average number 
of (un)employment insurance beneficiaries across CMA/CA for all 
years is about 125,000 and the average number of persons reporting 
federal child benefits is about 24,500 across CMA/CA and year.%
\footnote{%
The number of EI beneficiaries and the number of those collecting 
federal child benefits is higher on average when taking the 
subsample of CMAs only as shown in Table \ref{tab:sstats_cma}. 
This is because CMAs are much larger than CAs. 
} %
Table \ref{tab:sstats_cmaca} shows the average total number of 
licenses being approximately 10,000 per hundred thousand with 
approximately four PALs for every one RPAL in CMAs and CAs. Roughly
1 in 10 Canadians had a license. The CMAs only Table \ref{tab:sstats_cma} 
shows a relationship closer to seven PALs for every two RPALs, implying a 
lower share of restricted licenses among total licenses in CAs relative to
CMAs. Similarly, the CMA only sample has a lower concentration 
of total registered and restricted firearms (2526 per hundred 
thousand on average) than the CMAs and CAs sample (3658 per hundred 
thousand on average). In both geographies, handguns make up the 
lion's share of the registered and restricted firearms, 
representing about eight-ninths of the total guns in CMAs and 
CAs and about 90\% in CMAs only. To summarize, our less-urban 
sample is more likely to have higher concentrations of restricted 
licenses, restricted firearms, and a greater rifle-to-handgun ratio 
on average. The remaining registered and restricted firearms in 
our data are approximately two orders of magnitude fewer than 
the number of registered and restricted rifles. It is important to note that 
these figures understate the true number of firearms in Canada, as not all legally 
owned guns are required to be registered and some are owned illegally. However, in 
contrast to the United States, where gun ownership is significantly higher and 
unregistered ownership is widespread, Canada's firearm registration system provides 
a more limited but structured view of gun prevalence. Finally, Table
\ref{tab:sstats_cma} shows that firearms crimes, homicides, 
and murders per hundred thousand are quite close to zero.  

%%%%%%%%%%%%%%%%%%%%%%%%%%%%%%%%%%%%%%%%%%%%%%%%%%%%%%%%%%%%%%%%%%%%
\subsection{Model and Empirical Strategy} \label{section:strategy}

We are interested in recovering the causal effect of firearms on 
outcomes like deaths and crimes. While other 
papers make use of more modern techniques in causal inference 
\citep{Card2022, Imbens2022} such as difference-in-differences in 
\citet{Donohue2022} and \citet{Ferrazares2022}, causal effects in this paper 
rely on strict exogeneity between our firearms measures 
and our outcomes after conditioning on unobserved and fixed effects at the 
geographic level \citep{Wooldridge2010}. These assumptions are similar to papers
in this literature like \citet{Duggan2001}, \citet{Cook2006} 
and \citet{Khalil2017}, who also rely on variation in firearms 
ownership across time and geography.\footnote{The implicit assumption
made in these papers (using \citet{Wooldridge2010}'s naming convention) 
is consistent with the strict exogeneity assumption between firearms measures and 
outcomes once conditioning on the unobserved (fixed) effect at their geographic 
level. This allows for lags of both outcomes and covariates to have an effect 
on the contemporaneous outcomes and firearms measures.}  Our identifying assumption
is strict exogeneity of the firearm measures conditional on fixed effects and 
observed covariates. In particular, we assume that the idiosyncratic error term
is mean independent of the full history of firearm measures and covariates. This 
assumption rules out feedback from past outcome shocks into current firearm measures
after conditioning on fixed effects and controls, ensuring consistency of 
the fixed effects estimator.

A concern is that there exists a series of observable 
factors which affect: (1) the likelihood to purchase firearms and 
(2) firearms-related deaths. Examples include local economic conditions 
expressed in an unemployment rate or access to financial supports 
as well as the differences in socio-economic variables like 
the proportion of females (who are less likely to own firearms). 
We use similar variables in our analysis, denoting them by 
$\mathbf{x}$ to partially address concerns regarding omitted 
variables.

There also likely exist omitted variables which are not 
easily observed and cannot be proxied for, 
including heterogeneity across provinces ($\pi$), CMA/CA ($\xi$) 
and time ($\tau$). A motivating example for including
geography fixed effects would be places' different
preferences or their underlying desire to own firearms as well 
as places' likelihood for firearms to be used in deaths or crimes.
In so far as observed variables only partially correlate with 
these hard-to-observe variables, we can control for 
geography-specific unobserved variables using fixed effects.  
Additionally, differences across time in the landscape for owning 
firearms and using their associated deaths nationally would want 
to be controlled for. An example of this would be the implementation
of the \textit{Common Sense Firearms Act} in 2015 which grandfathered 
Possession and Only Licenses (POLs) into Possession and Acquisition
Licenses (PALs). Differences in the amenity value of firearms are important 
to control for. Unrestricted firearms can broadly be used on Crown and rural 
private land. Restricted rifles are only allowed in licensed ranges, often 
outdoors in rural areas. Handguns are allowed only in ranges, with indoor
ranges often being in urban areas.

Still, concerns remain regarding the use of firearms for crimes or
deaths. We display the typical concern in Appendix Figure \ref{DAG:usual} which 
shows an individual's proclivity for violence affecting gun ownership (and use) 
and the likelihood of committing crime in a Directed Acyclic Graph (DAG). 
Here, we abstract from the time dimension for clarity in our DAG, 
noting that they can be drawn with our strict exogeneity
between outcomes and firearms variables conditional on our fixed effects, 
as in \citet{Cunningham2021} and \citet{Kim2019}. 
Ideally (\textit{i.e.} holding constant a person's proclivity for
violence) one would be able to recover the causal effect of guns on
crimes. If this is relatively unchanging over time, place-specific 
fixed effects help to reduce this concern. However, the difficulty 
in measuring this proclivity for violence makes it hard to control for.  

We believe the relationship looks more like Appendix Figure \ref{DAG:ours_crime}. 
In this diagram, illegal firearms and legal firearms are unconditionally
unrelated, with no arrow drawn between the two of them. Proclivity for
violence affects illegal market guns and crimes. We draw a dashed red
line, with a blue `X' through it, running from proclivity for violence 
and into our legal guns variables. We are using a dashed, red line to suggest
an unlikely relationship, and a blue `X' through it to suggest no relationship,
between proclivity for violence and legal firearms. In addition to the 
background checks and training, license holders are subject to daily 
comparisons against their criminal record. Criminal convictions need 
to be disclosed  when applying for firearms licenses. This makes it quite
difficult for (violent) criminals to get a license. Finally, firearms are 
traceable to the people who purchase them, independent of whether the firearms
themselves are registered. This suggests that individuals wanting to use 
firearms for crimes would be best \textit{not} to use legal firearms. 

At the same time, the relationship between legal firearms and 
crimes is not always clear. \citet{Khalil2017} finds crimes are most 
likely being driven by illicit market guns and not by legally owned 
firearms in the United States. Also in the United States, 
\citet{Ferrazares2022} find (modest) voluntary firearms buybacks are unrelated 
to crimes in a city. On the contrary, 
\citet{Donohue2022} find that right-to-carry laws increase crime.  

The relationship between firearms and self-harm is depicted in the 
directed acyclic graph in Appendix Figure \ref{Figure:DAG_selfharm}. Unlike
the case for the effects of legal firearms on crimes, we believe 
depression is more likely to impact the use of legal guns for 
self-harm. While the licensing procedure attempts to screen out people who 
have a history of self-harm and provides phone numbers for those 
worried a legal firearms owner may hurt themselves, the episodic 
nature of depression makes it hard to completely screen out \cite{mortimer2020guns}.

Moreover, firearms owners may have been less affected by depression when 
applying to own a firearm than when they may be susceptible to
self-harm. Supposing the depiction in Appendix Figure \ref{Figure:DAG_selfharm},
together with a positive correlation between legal firearms ownership and depression
\citep{Perlis2022}, implies that our coefficient estimates would overestimate the true
relationship between legal firearms and self-harm if depression independently increases
the risk of self-harm. Evidence of a positive relationship between firearms and self-harm
\citep{Leigh2010} informs our prior on the likely sign of the true effect, but does not 
determine the direction of the bias itself. On the one hand, this would make us more 
likely to find a positive and statistically significant result. On the other hand, 
failure to detect an effect is suggestive that a relationship may not 
exist at all.

%%%%%%%%%%%%%%%%%%%%%%%%%%%%%%%%%%%%%%%%%%%%%%%%%%%%%%%%%%%%%%%%%%%%
\subsubsection*{CMA and CA Analysis}

We model the relationship between gun ownership and various outcome
variables using Equation \ref{eqn:Main}:
\begin{equation}\label{eqn:Main}
\text{Outcome}_{pct}= \ \beta \ \text{Guns Variables}_{pct} 
 	\ + \ \mathbf{x}^{\text{T}}_{pct} \gamma 
 	\ + \ \pi_{p} 
 	\ + \ \xi_{c} 
 	\ + \ \tau_{t} 
 	\ + \ \varepsilon_{pct}
\end{equation}
Observations are indexed by their province (subscript $p$), CMA/CA 
($c$), and year ($t$).

$\text{Guns Variables}_{pct}$ represents our variable of interest in 
all of our models and takes on values such as the number of PALs, 
the number of RPALs, the total registered guns, and the total type 
of registered and restricted guns, all per hundred thousand. 
In some instances, we include multiple guns variables 
in a given model. %
The variable $\text{Outcome}_{pct}$ represents a generic dependent variable 
which  changes based on the outcome we are modelling. Outcome variables
include the total number of deaths, the total number of crimes, 
or the total number of a type of crime. %

$\mathbf{x}_{pct}$ is a vector of covariates that include the: 
average age, proportions of individuals who are female, the proportion 
of individuals who are married, the number of individuals claiming 
employment insurance, and the total number of 
individuals claiming federal child benefits. %
Province and CMA/CA fixed effects are represented by coefficients 
$\pi_{p}$ and $\xi_{c}$, respectively. Time fixed effects are 
represented by $\tau_{t}$. We interpret $\beta$ from our model as 
representing the marginal  effect between our 
$\text{Guns Variables}_{pct}$ and our dependent  variable, conditional 
on our covariates. %

In many scenarios coefficient estimates are small in magnitude. This
is partially a function of scaling but also because of the large values 
in our guns variables (usually between $10^1$ and $10^4$) and the small
values in our dependent variables (usually $10^1$). We often discuss 
estimates  which arise from $\frac{1}{\hat{\beta}_{x}} = \hat{\beta}_{x}^{-1}$ 
which we interpret as how many $x$ would be necessary to increase the outcome 
variable by $1$, \textit{ceteris paribus}. When looking at our crime data 
by CMA, we have opted to standardize our guns variables for ease
of interpretation and due to small coefficient estimates. %

Equation \ref{eqn:Main} uses residual variation from the sample average deviations 
within CMA/CA-province-years for both outcomes and guns variables. Some of the 
variation in firearms likely comes from people who choose not to renew their 
licenses and those who are first-time license holders. Additionally, variation
may come from changes in the ownership of registered and restricted firearms. 
Data identifying license renewals versus first-time owners, and changes in legal
firearms, represent ideal data which we do not have. Confiscated firearms or deaths
of gun owners may also result in legal supply of firearms fluctuations. What we hope
to identify is how many additional firearms deaths result from above-average 
deviations in license holders and legal firearms. 

In many cases, the true number of deaths from firearms (less so for crimes) is
quite small within any CMA/CA annually. Taking the deviations from that trend 
(CMA/CA fixed effects) further reduces the residual variation in deaths or crimes. 
The trade-off is using the between- versus within-variation to try and identify 
these effects. The former fails to control for unobserved heterogeneity, likely 
biasing estimates, while the latter has few observations for any CMA/CA, making 
it hard to detect true effects. Recognizing our setting, we present both of 
these results. We ultimately continue using the regressions containing geographic
fixed effects to better align ourselves with previous literature.%
\footnote{Future work can take advantage of the firearms data given to us at the 
census subdivision. Doing so would give us smaller regions of Canada which may
better divide urban and rural Canada. This may sufficiently allow us to control 
for common unobserved heterogeneity at the CMA, CA, or non-CMA/CA while having 
variation \textit{within} each region. This would simultaneously help with issues
associated with power as there are many CSDs in any one CMA, CA or non-CMA/CA. 
Even without controlling for region fixed effects, finer geographies would likely
reduce coefficient estimate because: (1) firearms deaths happen in low-population 
regions, and, (2) low-population places typically have many firearms.}

\section{Results} \label{section:results}

%%%%%%%%%%%%%%%%%%%%%%%%%%%%%%%%%%%%%%%%%%%%%%%%%%%%%%%%%%%%%%%%%%%%%%
\subsection{National and Provincial Trends}

\begin{figure}[H]
	\begin{center}
    \caption{Canadian Restricted Firearms \label{fig:can_guns}}          
    \includegraphics[width=\textwidth]{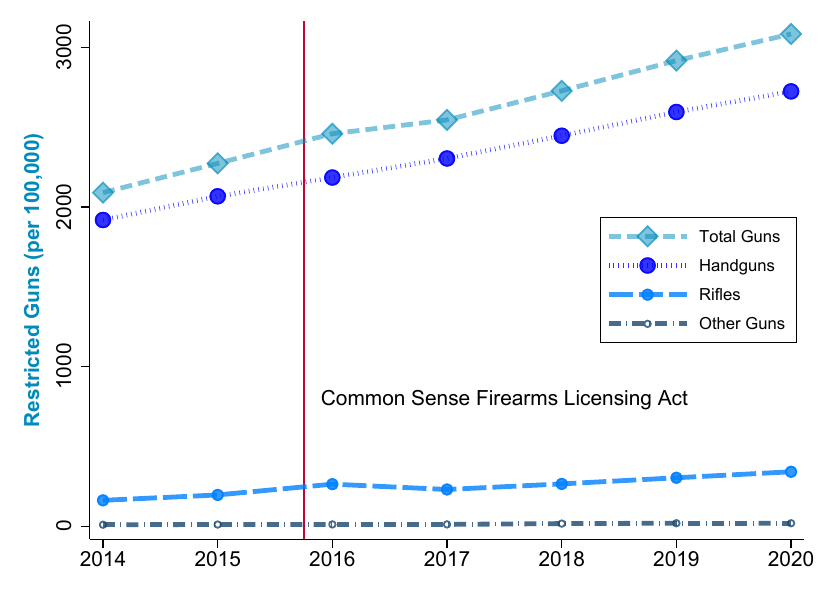}
    \end{center}
\footnotesize{
  Notes: Authors' calculations. %
  Guns data from \textit{Access to Information Request} from the %
  Royal Canadian Mounted Police (RCMP). %
  Scaled by population data from \citet{StatCan_PopChange}. %
} %
\end{figure}

\begin{figure}[H]
	\begin{center}
    \caption{Canadian Licenses \label{fig:can_licenses}}          
    \includegraphics[width=\textwidth]{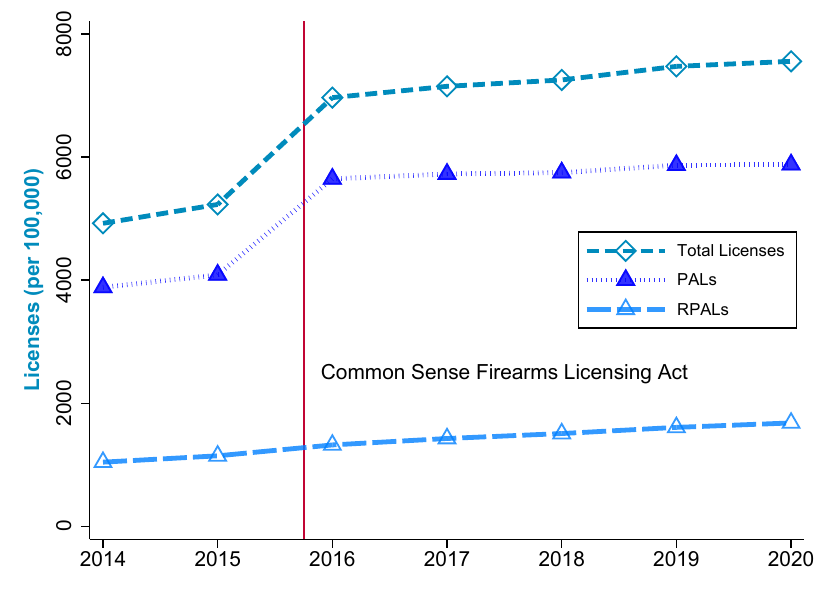}
    \end{center}
\footnotesize{
  Notes: Authors' calculations. %
  Guns data from \textit{Access to Information Request} from the %
  Royal Canadian Mounted Police (RCMP). %
  Scaled by population data from \citet{StatCan_PopChange}. % 
} %

\end{figure}

Figure \ref{fig:can_guns} and Figure \ref{fig:can_licenses} show the 
trends in national prevalence of restricted guns and licenses, respectively.
The data provided to us are as of December 31\textsuperscript{st} of each
year, which we correspond to the following year. Firearm stock measures are 
recorded as of December $31$ of each calendar year and are aligned with outcomes 
in the subsequent calendar year. Thus, end-of-year firearm prevalence for year $t$ 
corresponds to outcomes in year $t+1$. Figures reflect this alignment, which explains
why Figure 3 covers $2014-2020$ while the regression sample is described as 2013-2019
based on firearm measurement years. Figure \ref{fig:can_guns} 
shows the total restricted firearms which is the sum of three components: 
handguns, rifles, and other guns. The prevalence of restricted guns has been 
increasing between $2013$ and $2019$ primarily as a result of handguns. 
The prevalence of handguns has risen from just under $2,000$ at the end of
$2013$ to about $2,500$ at the end of $2019$. Rifles and other guns have 
remained relatively stable in our time period, and are significantly fewer 
than handguns. While there are hundreds of types of restricted rifles, the prevalence of 
other guns is near zero.  

Figure \ref{fig:can_licenses} shows the prevalence of firearms licenses has increased steadily over 
time, with a marked increase between $2015$ and $2016$ from about $5,000$ 
licenses to about $7,000$ licenses per $100,000$. After 2016, the relationship stabilizes
to about $7,000$ licenses with only small increases across our time period.  
The \textit{Common Sense Firearms Licensing Act} converted all Possession
Only Licenses (POLs) - licenses that allowed individuals to possess but not acquire 
new firearms - from the previous licensing system, to Possession and
Acquisition Licenses in September 2016, which we believe is the result of
the one-time increase in total licenses, as driven by PALs. 

\begin{figure}[H]
	\begin{center}
    \caption{Canadian Restricted Firearms, Licenses, Homicides and Suicides \label{fig:can_trends}}          
    \includegraphics[width=\textwidth]{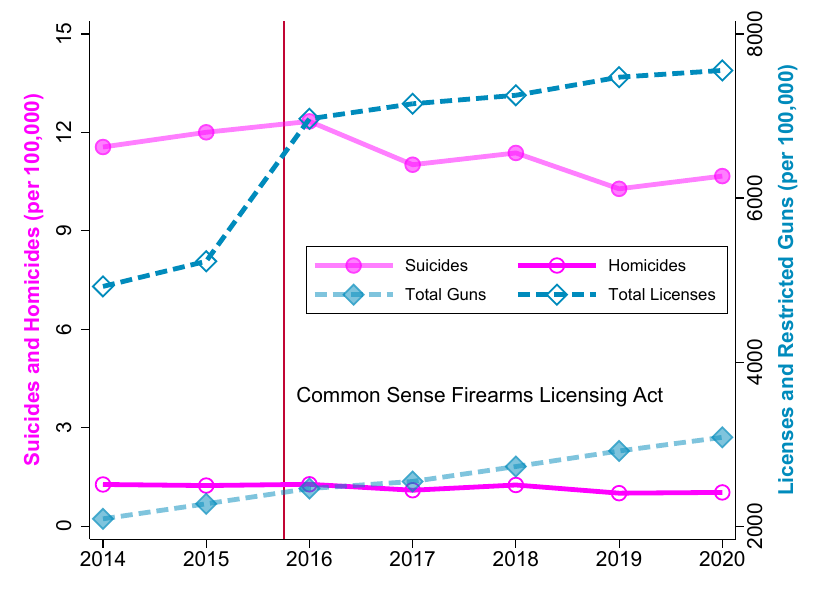}
    \end{center}
\footnotesize{
  Notes: Authors' calculations. %
  Guns data from \textit{Access to Information Request} from the %
  Royal Canadian Mounted Police (RCMP). % 
  Deaths data from \citet{StatCan_CauseDeath}. %
  Scaled by population data from \citet{StatCan_PopChange}. %
} %
\end{figure}

Figure \ref{fig:can_trends} shows national trends from 2014 to 2020 in suicides, homicides,
registered restricted firearms, and firearms licenses per 100,000 population. Suicide rates
remain relatively stable between 10 and just over 12 per 100,000, while homicides stay 
consistently below 2. Over the same period, both gun licenses and registered restricted 
firearms show a general upward trend.

We further characterize our data through choropleth maps which describe 
the Canadian spatial distribution of guns, licenses, homicides and suicides. The maps 
demonstrate a ranking of the census subdivisions by their rate per hundred thousand 
for these variables. All maps use values from $2019$ and each 
represents a different variable.

\begin{figure}[H]
	\begin{center}
    \caption{Distribution of Licenses and Restricted Firearms by Canadian Census Subdivisions during 2019 \label{fig:map_firearm}}
    \begin{subfigure}[b]{0.8\textwidth}
            \centering
			\caption{All Licenses (per 100,000)} \label{fig:map_license}            
            \includegraphics[width=\textwidth]{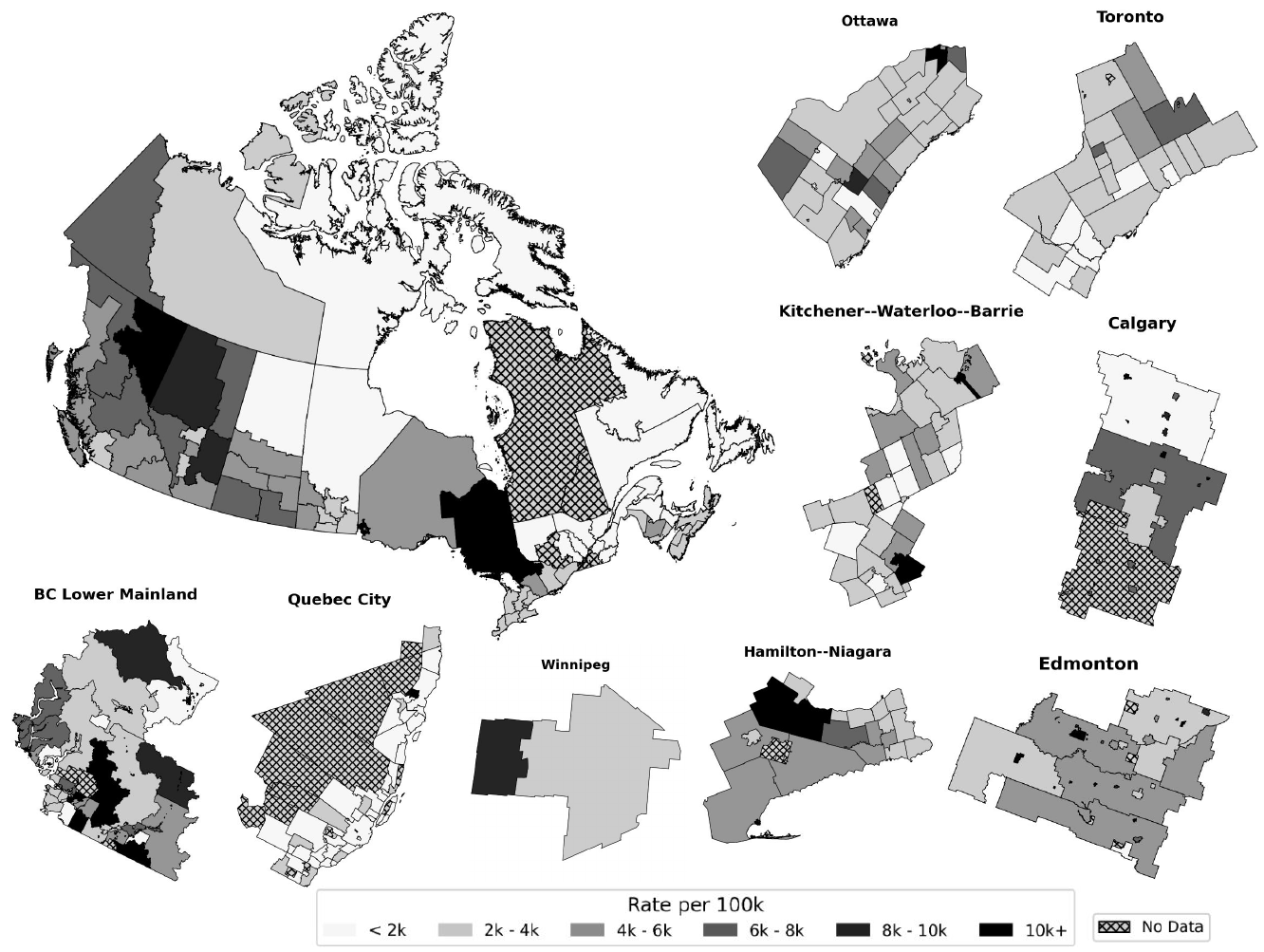}
    \end{subfigure}
    \\[1.0em]
    \begin{subfigure}[b]{0.8\textwidth}
            \centering
			\caption{Total Restricted Firearms (per 100,000)} \label{fig:map_all_guns}            
            \includegraphics[width=\textwidth]{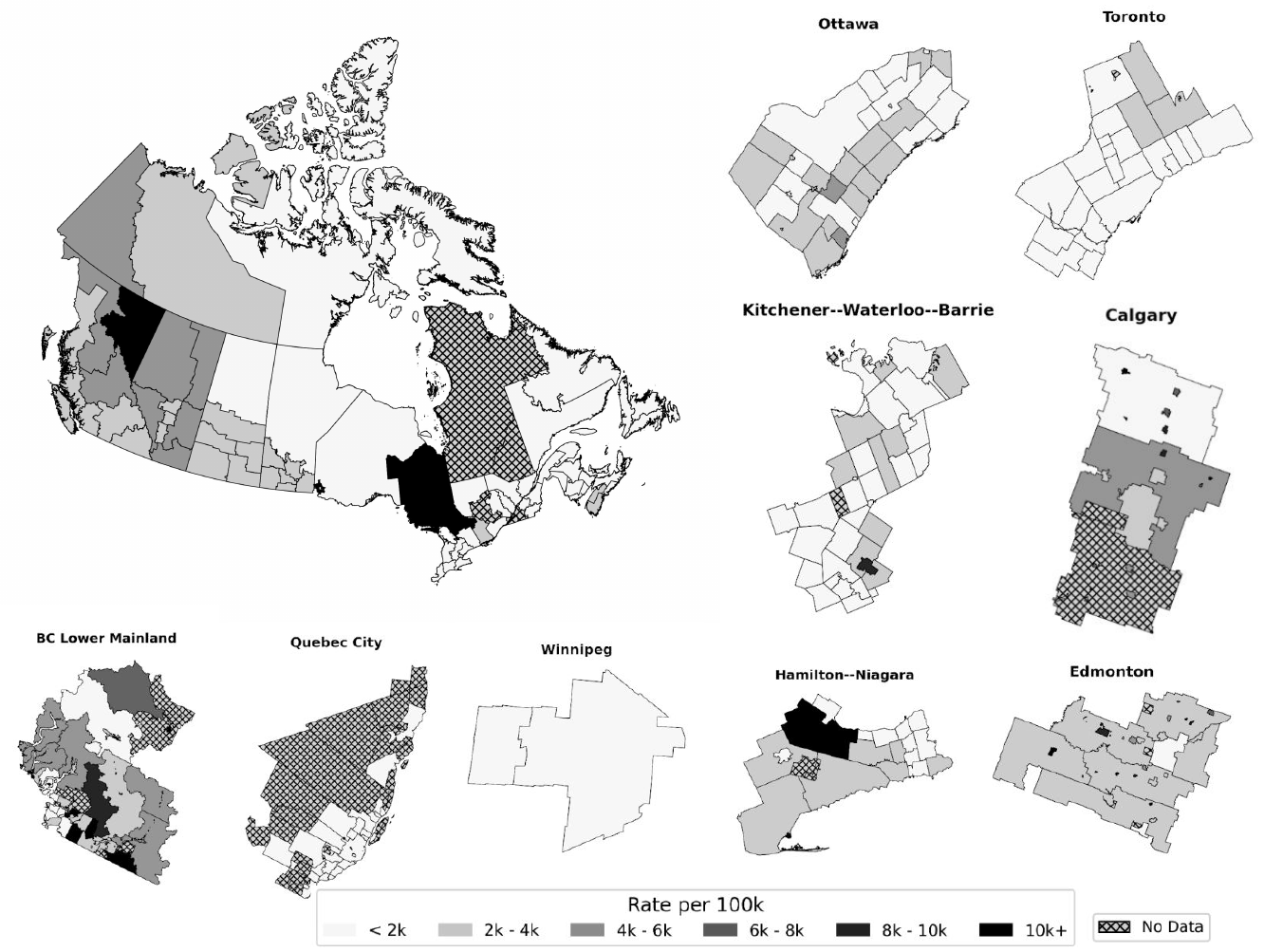}
    \end{subfigure}
    \end{center}
\footnotesize{
  Notes: Authors' calculations. %
  Guns data from \textit{Access to Information Request} from the %
  Royal Canadian Mounted Police (RCMP). % 
  Population data is from \citet{StatCan_PopChange}. %
  The map is constructed on the subset of data for 2019. 
  Additional zoomed in versions shown for major areas.%
  Panel A shows all licenses (PAL + RPAL) per 100,000. %
  Panel B shows all restricted firearms per 100,000. %
} %
\end{figure}

Figure \ref{fig:map_firearm} shows the spatial variation across 
Canada in 2019 for firearms licenses per hundred thousand population 
(Figure \ref{fig:map_license}) and the total registered restricted 
firearms per hundred thousand (Figure \ref{fig:map_all_guns}). There is significant geographic
variation in both licensing rates and the number of restricted firearms. Higher rates of firearm
licenses -- above $10,000$ per $100,000$ -- are observed in several northern and central regions of Canada, 
including Alberta, Ontario, and parts of British Columbia. In contrast, major urban areas such as Toronto
and Vancouver report much lower license rates, often under $4,000$. Similarly, the distribution of registered 
restricted firearms varies, with some subdivisions in Alberta, British Columbia, Ontario, and the
northern territories recording rates above $2,800$. Notably, certain regions, such as Hamilton-Niagara 
and the BC Lower Mainland, exhibit significant internal variation, with neighboring subdivisions differing
considerably in registered firearms. These patterns suggest that both geographic location and local context
strongly influence the distribution of firearm licenses and restricted firearms across the country.

\begin{figure}[H]
	\begin{center}
    \caption{Distribution of Restricted Rifles and Restricted Handguns by Canadian Census Subdivisions during 2019 \label{fig:map_type}}
    \begin{subfigure}[b]{0.8\textwidth}
            \centering
			\caption{Total Restricted Rifles (per 100,000)} \label{fig:map_rifle}             
            \includegraphics[width=\textwidth]{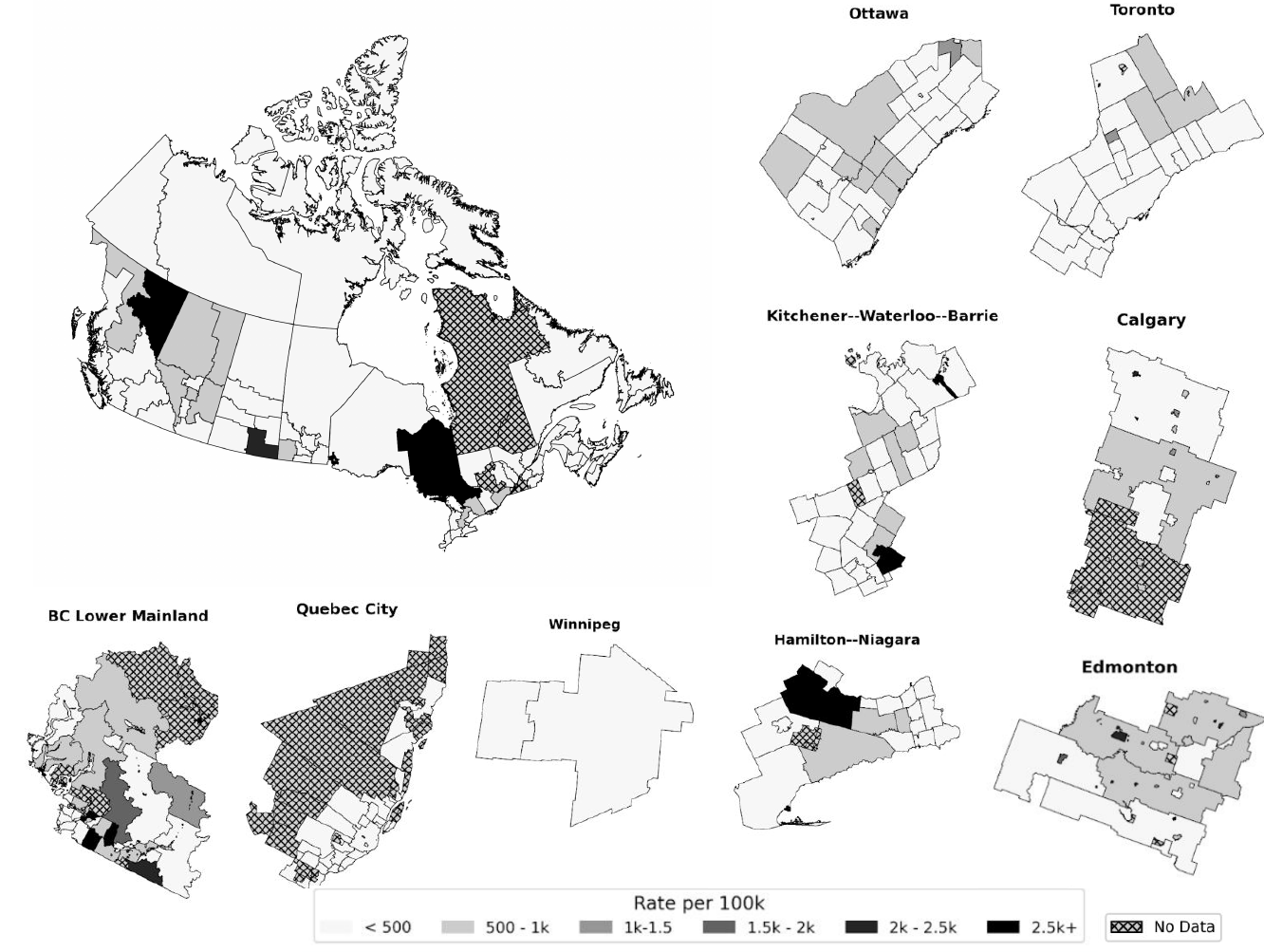}
    \end{subfigure}
    \\[1.0em]
    \begin{subfigure}[b]{0.8\textwidth}
            \centering
			\caption{Total Restricted Handguns (per 100,000)} \label{fig:map_handgun}             
            \includegraphics[width=\textwidth]{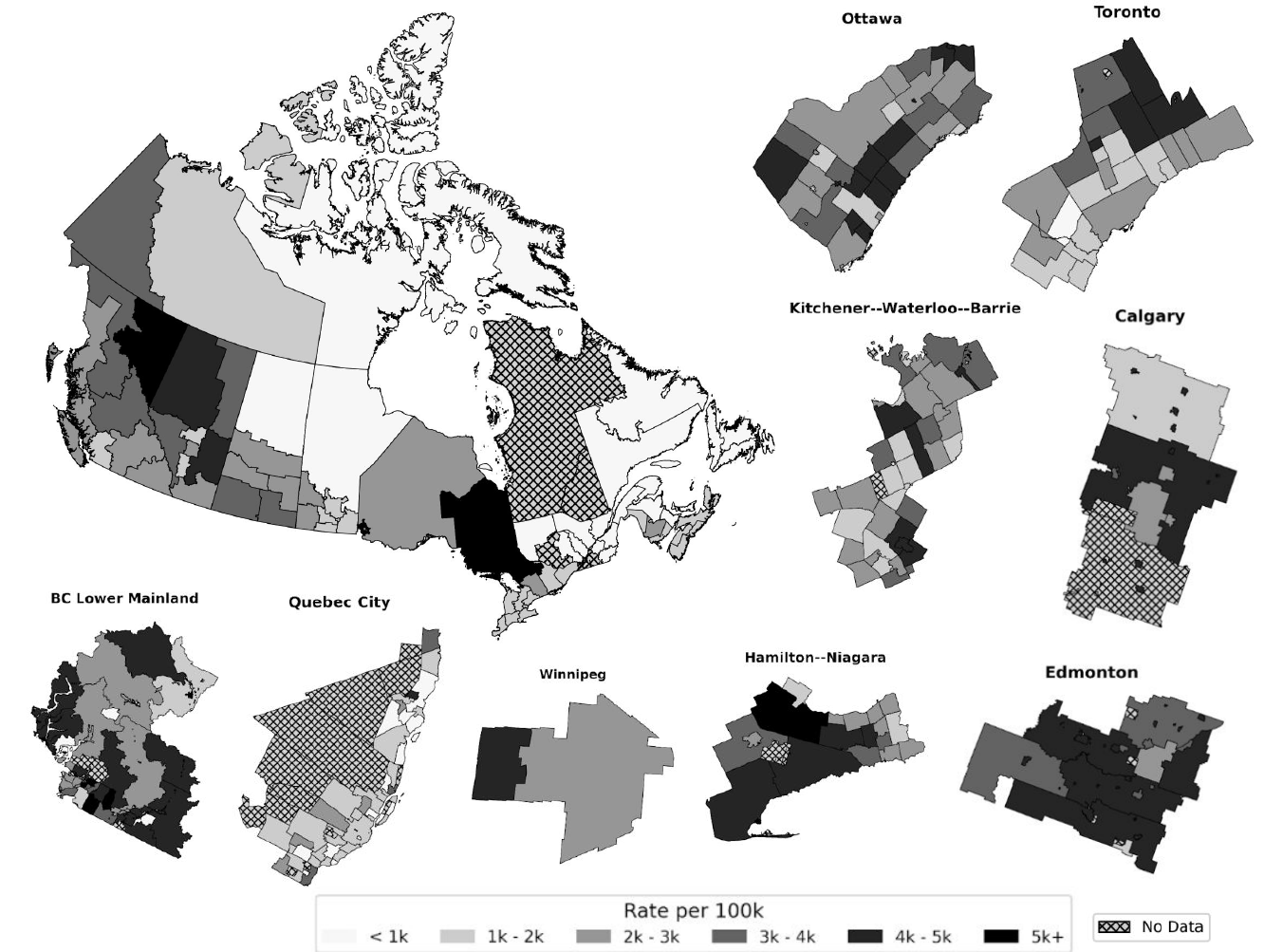}
    \end{subfigure}
    \end{center}
\footnotesize{
  Notes: Authors' calculations. %
  Guns data from \textit{Access to Information Request} from the %
  Royal Canadian Mounted Police (RCMP). % 
  All data is scaled by population data from \citet{StatCan_PopChange}. %
  The map is constructed on the subset of data for 2019. 
  Panel A shows total restricted rifles per 100,000. %
  Panel B shows total restricted handguns per 100,000. %  
  Additional zoomed in versions shown for major areas.%
} %
\end{figure}

Figure \ref{fig:map_type} shows the spatial variation across 
Canada in 2019 for total registered restricted rifles per hundred 
thousand (Figure \ref{fig:map_rifle}) and the total registered 
restricted handguns per hundred thousand (Figure 
\ref{fig:map_handgun}). Since these two figures represent a
breakdown of total of registered and restricted firearms, they closely
match Figure \ref{fig:map_all_guns}. The spatial pattern of restricted rifles (Figure \ref{fig:map_rifle}) 
highlights higher registration rates in parts of Alberta, Saskatchewan, rural Ontario, and northern British
Columbia, where some subdivisions exceed 2,500 rifles per 100,000. In contrast, restricted handguns 
(Figure \ref{fig:map_handgun}) are more concentrated in urban areas, particularly in and around Toronto, 
Hamilton, Edmonton, and the BC Lower Mainland, where some regions record over $5,000$ handguns per $100,000$. 
This divergence in spatial concentration illustrates a broader pattern: restricted rifles are more commonly 
registered in rural and northern regions, while handguns are more prevalent in urban areas. Together, these
maps emphasize the influence of the urban-rural divide and regional firearm use norms on the ownership patterns 
of different types of restricted firearms in Canada. One explanation for these differences is that urban 
areas often have indoor ranges which allow handguns but disallow rifles. Alternatively, more rural areas
have outdoor ranges which allow both types of firearms. 

\begin{figure}[H]
	\begin{center}
    \caption{Distribution of Homicides and Suicides by Canadian Province during 2019 \label{fig:prov_homSui}}
    \begin{subfigure}[b]{0.99\textwidth}
            \centering
			\caption{Total Homicides (per 100,000)} \label{fig:prov_homSui_A}             
            \includegraphics[width=\textwidth]{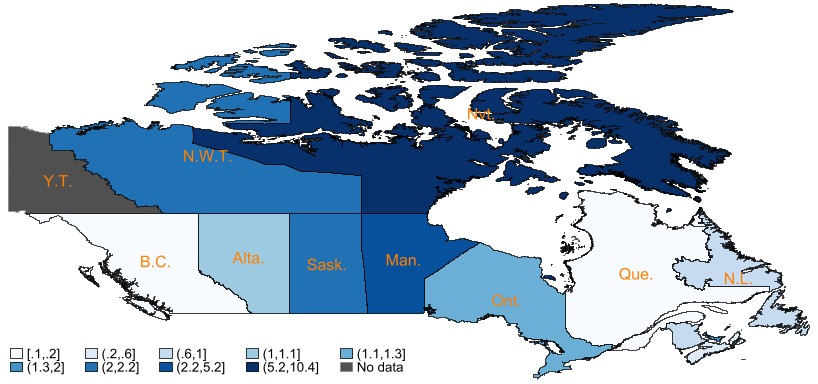}
    \end{subfigure}
    \\[1.0em]
    \begin{subfigure}[b]{0.99\textwidth}
            \centering
			\caption{Total Suicides (per 100,000)} \label{fig:prov_homSui_B}             
            \includegraphics[width=\textwidth]{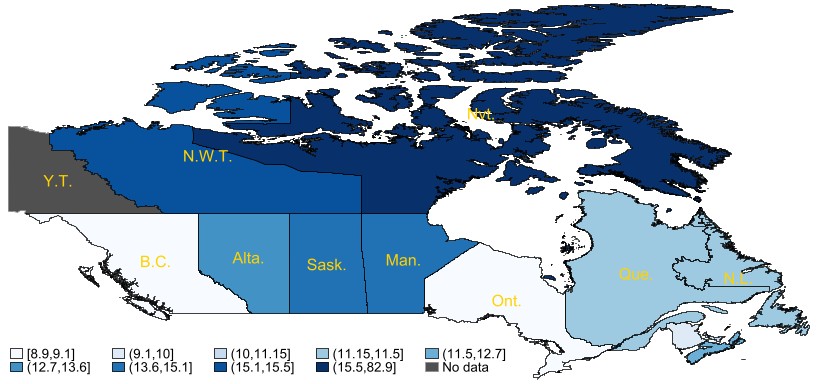}
    \end{subfigure}
    \end{center}
\footnotesize{
  Notes: Authors' calculations. %
  Deaths data from \citet{StatCan_CauseDeath}. %
  All data is scaled by population data from \citet{StatCan_PopChange}. %
  The map is constructed on the subset of data for 2019.
  Panel A shows total homicides per 100,000. %
  Panel B shows total suicides per 100,000. %   
  New Brunswick, Nova Scotia, and PEI are not labelled %
  (bottom-right corner). %
} %
\end{figure}

Figure \ref{fig:prov_homSui} shows the spatial variation across provinces 
in 2019 for homicides (Figure \ref{fig:prov_homSui_A}) and the total 
suicides per hundred thousand (Figure \ref{fig:prov_homSui_B}). 
In general, the rankings between these two variables are quite different. 
Homicides are relatively more concentrated in Nova Scotia, New Brunswick, 
Ontario, Manitoba and Nunavut. Suicides are more concentrated in New Brunswick, 
Manitoba, Saskatchewan, Nunavut and Northwest Territories. 

Next we explore the relationship across provinces over time using pooled OLS regressions 
(i.e., without province fixed effects). Table \ref{tab:prov_crimes} shows the coefficient estimates 
when modelling provincial level gun crimes as a linear 
function of licenses (or total registered firearms) per hundred 
thousand and that variable's annual percentage change.%
\footnote{We model the relationship between various outcomes variables and our guns 
variables (total firearms licenses or total restricted and registered firearms) (per $100,000$) 
at the provincial level as:
\begin{equation}\label{eqn:Prov}
y_{pt} = 
    \ \beta_{0} 
    \ + \ \beta_{1} \ \text{Guns Variables}_{pt} 
    \ + \ \beta_{2} \ \% \Delta \text{Guns Variables}_{pt} 
    \ + \ \varepsilon_{pt}.
\end{equation}}
 Eight columns 
of gun crimes are considered in Table \ref{tab:prov_crimes} including 
total firearms violations, discharging a firearm, using a firearm with 
intent, firearms documentation and administration (labelled as 
``documentation issue'' in the table), improper storage 
of firearms, homicides, first-degree murder, second-degree murder, and 
manslaughter. The top panel shows when our main independent variable is 
total licenses (PALs plus RPALs). %
With the exception of firearms documentation offenses and first-degree murders, 
which show a negative and no relationship with total licenses respectively, the 
remaining crime outcomes exhibit a positive relationship with total licenses, 
although most of these associations are modest in size.%
\footnote{For example, even after multiplying the total licenses coefficient estimate 
by 1000 (slightly less than a one-eighth increase relative to the mean) 
the total number of firearms violations would increase by 0.128 
per hundred thousand. This is practically small, considering the average 
number of total firearms violations is 9.4 per hundred thousand, such a 
small response in our dependent variable relative to a large hypothetical 
increase in our independent variable. There appears to be no relationship 
between the annual percentage change in total licenses and all columns 
of our top panel.} %
The bottom panel shows the total registered and restricted firearms as 
the main independent variable. In this case, all columns show a positive 
relationship between the total level of registered and restricted 
firearms and crime outcomes, excepting the null relationship in the 
firearms documentations column. Again, there appears to be no relationship 
between annual percentage changes in registered and restricted firearms. 
Finally, the practical significance of these coefficient estimates is small 
for analogous reasoning to the top panel: the level effect necessary to 
generate a large dependent variable response is unrealistically large. 

Table \ref{tab:prov_ols} shows the coefficient estimate when modelling
provincial level deaths associated with suicides, homicides, accidents, 
and all other deaths, with total licenses (top panel) and total registered 
and restricted firearms (bottom panel) as in Equation \ref{eqn:Prov} and 
akin to the previous table. Columns 1-4 use the full sample (both sexes), 
while Columns 5-8 and 9-12 are restricted to males and females, respectively. 
Across both panels, and excepting 
the residual ``other'' deaths category which is negative and statistically 
significant for both sexes and only males, annual percentage changes 
are not associated with deaths across any of the samples. The level of licenses 
coefficient estimates are positive and statistically significant for: suicides 
across all samples; accidental deaths across males only (with a negative coefficient
for females); and all other deaths across both sexes and males only. The total number of 
registered and restricted firearms coefficient estimates of the levels 
are associated with positive and significant effects for homicides across 
all samples, accidents for both sexes and only males samples, and accidents 
for the only males samples. Similar to the previous table, even large 
changes in the levels of restricted guns or licenses (\textit{i.e.} 
multiplying our coefficient estimate by 1000) would be associated 
with very small changes in deaths. 

We draw three conclusions suggested from the previous figures and tables. 
First, homicide and suicide rates remain relatively stable between 2013 and 2019, 
whereas total licenses and total registered and restricted firearms increase steadily
over this period. When looking at 2019 only, the ranking of provinces for homicides, 
suicides, total licenses and registered and restricted firearms is not 
invariant: the choropleth map colours change with each of the different 
measures. Second, when modelling the relationship between crimes (or deaths) 
and our guns variables, levels are more likely to matter than annual 
percentage changes. Third, any of the effects which we might be picking up 
and which may appear to align with the intuition that guns are associated 
with more crimes and more deaths, are economically insignificant due to the size of 
the coefficient estimates. This last point should be understood relative to 
the model, estimation and data, all of which are simple and exploratory and 
mask substantial heterogeneity. To better understand potential relationships, 
our next section explores models aimed at similar relationships at 
finer geographic levels. 

\subsection{Deaths Due to Firearms}

Firearms-related deaths are a small fraction of total deaths in 
Canada, even conditional on age and 
sex, suggesting our previous measure is unlikely to be able 
to detect a signal from the noise. We attempt to overcome this data limitation 
by accessing administrative deaths data. 

Table \ref{tab:rdc_all_deaths} uses the deaths where a firearm 
was the leading cause as a dependent variable following Equation \ref{eqn:Main}.
Estimates use ordinary least squares, weighting
all observations with their annual population and clustering standard 
errors at the provincial level. It should be noted that due to the small number of 
unbalanced clusters, these standard errors are biased downwards \citep{MNW-guide}. 
Columns represent changes in the included control variables, while panels vary the
gun-related variables being examined.  

Panel 1 shows the coefficient estimates across different specifications only 
including restricted firearms as the guns variable. Panel 2 only 
has licenses as our guns variable. Panel 3 uses both guns and 
licenses. Panel 4 breaks our licenses into PALs and RPALs while 
including total restricted guns. Panel 5 breaks out our guns 
variable into three categories: handguns, rifles, and ``other guns.'' 
whilst including PALs and RPALs.  The other guns
category includes the other firearms from earlier: restricted 
commercial versions, shotguns, machine guns and submachine guns. 

Panels 1, 3, and 4 of Table \ref{tab:rdc_all_deaths} show no statistically significant effects of 
total restricted firearms on the total number of firearms deaths 
across every model. Panels 2, 3, and 4, show positive and statistically 
significant coefficient estimates of granted licenses on total firearms 
deaths in columns 1 through 4. Coefficient magnitudes are approximately
$0.29$ for all licenses, suggesting about $3,330$ fewer license 
holders (per $100,000$) would reduce one firearms death (per $100,000$). 
Similar effect sizes can be 
seen in Panel 5 for our two license types in columns 1 through 4 when
also controlling for handguns, rifles, and other guns. Rifles and the neither rifles nor 
handguns category are not significantly different from zero in 
any of the columns 1 through 4. Our handguns coefficient in column 4 of 
panel 5 is $0.12$ and significant at the $5\%$ level. This suggests that $8,333$ fewer  
handguns (per $100,000$) would reduce one firearms death (per $100,000$). 

Column 5 of Table \ref{tab:rdc_all_deaths}
introduces CMA and CA fixed effects. Coefficient estimates of licenses 
are no longer statistically significant in column 5 of any of our panels and 
also decrease by an order of magnitude. The neither handguns 
nor rifles category is the only statistically significant variable in column 5, 
equalling $1.97$. This suggests that $508$ fewer license holders (per $100,000$) 
would reduce one firearms death (per $100,000$). 

We choose to follow the specification in column 5 of Table \ref{tab:rdc_all_deaths} 
in our successive models for two reasons. First, the lack of parameter stability between 
columns 4 and 5 suggests there is considerable unobserved heterogeneity 
across CMAs and CAs. Column 5 takes this into account in the estimation 
whereas column 4 omits these fixed effects. This also represents a key 
advantage of panel data: to control for unobserved heterogeneity, which 
often includes places' time-invariant characteristics, like desire for
firearms ownership and total deaths from firearms.  
Second, \citet{Cook2006}, \citet{Leigh2010}, and 
\citet{Khalil2017}, include their finest geographic fixed effect in \textit{all} 
of their models. In \citet{Duggan2001}, all models are over differences in dependent 
and independent variables, making fixed effects inestimable. This helps 
us remain consistent with a literature that believes in controlling for place-specific
fixed effects.   

% TABLE FOR INTENT OF FIREARMS DEATHS
One question is whether the intent surrounding someone's death 
matters for our legal licenses and firearms variables. This 
is especially important if assaults or suicides are more likely to 
be correlated with legal firearms. Table \ref{tab:rdc_deaths_intent} 
shows the prevalence of firearms deaths by intent. There are three 
columns representing the three different types of deaths by intent: 
assaults, self-harm (interpreted as suicides) and unknown. 
We present three panels which represent three different models for our 
guns variables: panel 1 models total licenses and total restricted 
firearms jointly; panel 2 includes total restricted guns, PALs, 
and RPALs, while panel 3 has handguns, rifles, other guns,
PALs and RPALs. All columns control for year, province, CMA/CA fixed effects
as well as time-varying covariates.

Firearms deaths whose intent is considered to be assaults, self-harm, 
or unknown, are statistically insignificant across all three models at all 
traditional levels with respect to our licenses variables. This suggests 
that being able to legally acquire a firearm may not affect deaths.  Panel 1 
suggests total restricted firearms do not affect the assaults, suicides 
or deaths whose intent is unknown.\footnote{When modelling either total
licenses or restricted firearms in models where they are the only key
variable of interest, all results are also statistically insignificant
at all traditional levels.}
Total restricted firearms in a CMA/CA do not appear to affect deaths 
by assaults or whose intent is unknown (panel 2). Total restricted
firearms has a coefficient equaling $-22.86$ and statistically significant 
at the $5\%$ level with respect to self-harm (panel 2, column 2). However, 
we fail to reject an $F$-test for joint significance for all of the firearms 
variables in panel 2, suggesting they add little explanatory power.  

We decompose the total restricted guns in Panel 3 into handguns, rifles and other guns.
Handguns are statistically insignificant, while the rifles and other guns coefficients 
in the self-harm specification are statistically significant, with estimates of $-21.68$ 
and $1273.45$, respectively. The magnitude of the rifles coefficient is similar to the total
restricted guns effect in the previous panel. However, the other guns 
coefficient is two orders of magnitude larger, suggesting a reduction 
of these firearms by $787$ (per $100,000$) may decrease one death (per $100,000$). 
Our takeaway from these results, in conjunction with table \ref{tab:rdc_all_deaths} 
is that all firearms deaths are most affected by our other guns 
category, which appears to be coming from suicides, once taking into 
account our unobserved heterogeneity at the CMA/CA level. Although these firearms are not 
substitutes for practical use, since restricted firearms such as rifles can only be legally
used at licensed ranges, their statistical association with firearm deaths may reflect 
underlying risk factors or accessibility patterns.

%TABLE FOR DIFFERENT TYPES OF FIREARMS DEATHS 
An additional question is how well our legal firearms variables, and 
especially the \textit{type} of legal firearms, correspond to the 
type of firearm suspected to be the cause of death. If one supposes
that all deaths by handguns are attributable to \textit{legal} handguns, 
then we should expect positive and statistically significant coefficient 
estimates on our handguns variables. 

Table \ref{tab:rdc_deaths_types} shows the different types of
firearms involved in the firearms deaths. Handgun deaths, rifle deaths, 
unspecified firearms deaths and uncategorized firearms deaths, are shown
in columns 1, 2, 3, and 4, respectively. Panels 1, 2 and 3 are the same models
as Table \ref{tab:rdc_deaths_intent}. Again, when modelling total restricted 
firearms and total licenses, we fail to detect effects at all traditional 
levels. We see that restricted firearms 
effects on handgun deaths are positive and statistically significant in 
panels 2, and 3, with coefficient estimates of $5.58$ (approximately $179$ additional firearms 
associated with one additional death per 100,000 people) and $17.97$ (approximately $56$  additional 
firearms associated with one additional death per 100,000 people), 
respectively. In column 1 of panel 3, we also see that rifles are positive and
statistically significant with respect to handgun deaths. While an order of
magnitude less than handgun effects, it is somewhat surprising to be statistically 
significant. In any case, the amount of firearms necessary to change the amount of firearms-related 
deaths are well beyond what most CMAs or CAs have as their stock. Licenses variables and the neither
rifles nor handguns category are not significant in column 1. 

Column 2 of panel 2 shows restricted firearms effects 
on rifle deaths are negative and statistically significant, suggesting rifle 
deaths decrease with additional restricted firearms. This effect is shown to 
be coming from rifles and neither rifles nor handguns in panel 3, both of which 
are negative and statistically different from zero.

Column 3 of Table \ref{tab:rdc_deaths_types} shows the prevalence of
firearms deaths where the firearm type is unspecified. Panel 2 and 3 show 
possession and acquisition licenses (PALs) coefficient estimates are 
$190.92$ (implying roughly $5.2$ additional licenses associated with one additional 
death per 100,000 people) for either model, and statistically significant at the $10\%$ 
level. Total restricted firearms is statistically indistinguishable from
zero (panel 2) but when broken down into its components, the other guns 
coefficient equals $2467.60$ (approximately $0.4$ additional firearms associated with one 
additional death per 100,000 people) and statistically significant at the
$1\%$ level. Uncategorised firearms deaths appear unaffected by our firearms 
variables in either model. 

Between Tables \ref{tab:rdc_all_deaths}, \ref{tab:rdc_deaths_intent} and 
\ref{tab:rdc_deaths_types}, a clearer relationship between 
our legal firearms variables and firearms deaths is emerging. First, 
being able to acquire firearms (\textit{i.e.} our licenses variables) 
are almost always unrelated to our firearms deaths variables once
taking into account CMA/CA unobserved heterogeneity.  Again, this is interesting
as we saw a 70\% increase in restricted licences over the period. Moreover, total firearms 
deaths, firearms deaths whose intent was assaults and self-harm, as well as 
deaths with unspecified firearms appear to be positively related to our 
residual firearms category (including restricted shotguns, submachine guns, 
commercial versions and machine guns). 

Less clear are the effects of specific legal guns on firearms deaths 
specific to a type of firearm, like a handgun or rifle. While handgun deaths are positively
related to handguns, rifle deaths seem to be negatively correlated with 
legal rifle ownership. The effects on rifles run counter to our intuition for legal firearms
to be a driver of firearms deaths. One 
possible explanation is that the majority of rifles are unrestricted 
and hence, do not enter our counts of \textit{restricted} firearms.\footnote{At the end of 
the long gun registry period the Canadian government reported that  ``approximately 7.1 million 
of the total 7.9 million (90\%) firearms registered in Canada were non-restricted.'' 
\citep{CanadaGazette2021_LongGunRegistry}} Unrestricted rifles require access to
private (or crown) land, outside of municipalities to use, whereas restricted rifles
require the population density to support a licensed range. There may exist a 
negative correlation between restricted rifles and unrestricted rifles, 
especially if they are substitutes in use.  

This would also justify the
statistically significant and negative coefficient on the neither rifles
nor handguns category in the rifle-death specification (Table 8), since they are 
composed of guns most likely to be long-guns (types of shotguns, commercial 
versions, submachine guns and machine guns).

What the evidence suggests is that had fewer firearm licenses been granted or had the 
overall number of firearms been lower, we would not necessarily observe a universal 
reduction in total firearm-related deaths.

%%%%%%%%%%%%%%%%%%%%%%%%%%%%%%%%%%%%%%%%%%%%%%%%%%%%%%%%%%%%%%%%%%%%%%
\subsection{CMA-level Crimes}

Our deaths data, in either dataset, do not have what type of crime the 
deaths were classified as. They also only pick up fatality costs of 
firearms harms while missing outcomes on society not involving death.  
Data pertaining to firearms-related crime, homicides, murders and manslaughter
are readily available at the CMA level outside a Data Centre. 
While this restricts our total number of observations, it allows us to 
investigate outcomes which are publicly available.

Table \ref{tab:cma_firecrimes} shows the coefficient estimate when 
modelling CMA level crimes associated with firearms violations according
to Equation \ref{eqn:Main} and our preferred specification estimated 
using ordinary least squares. Noting our previous small coefficient 
estimates, we standardize our guns variables for easier interpretation. 
Panels one through five vary the main 
independent variables involved in the model. Columns one through nine 
vary the dependent variable and denote total firearms violations, 
discharging a firearm with intent, using firearms in commission of an 
offence, firearms documentation and administration issues, unsafe 
firearms storage, homicides, first-degree murder, second-degree murder, 
and manslaughter. Panel one uses the prevalence 
of all restricted guns. Panel two uses the prevalence of all licenses. 
Panel three is a model with both total firearms and total licenses.

We focus our attention on panel three since coefficient estimates 
in panel three resemble the results from panels one and two: coefficient 
estimates have the same sign and approximately the same values when 
modelling guns and licenses separately (due to the correlation between 
licenses and guns). Additionally, panel three helps contextualize the 
results in panels four and five which successively disaggregate licenses 
(panel four) then firearms (panel five) by their respective types. Panels four and 
five build on the results from Panel three by refining the coefficient estimates 
and providing greater clarity on which firearm variables are associated with 
specific crime outcomes. Panel three suggests no effects of restricted 
firearms on any of  our crime outcomes. Total licenses has no effects on
total firearms  crimes or crimes related to firearms discharges,
using firearms,  improper storage, or manslaughter. A one standard
deviation increase in the number of licenses in a CMA would decrease 
the prevalence of documentation issues by about $1.56$ per annum. We see 
that the prevalence of homicides is decreasing by approximately 
$0.332$ for a one standard deviation increase in the number of licenses. 
This appears to be a result of two opposing effects: increasing 1st
degree murders (coefficient estimate being $0.529$, significant at the
$10\%$ level) and decreasing 2nd degree murders (coefficient estimate 
being -0.883 and significant at the 5\% level).

Panel four models our key explanatory variables as total restricted guns 
and breaks out licenses 
into PALs and RPALs. In this way, we are seeing if it is the \textit{type}
of license owner which may be causing some of our changes in crimes. 
Columns one through six, all non-homicides as well as all homicides, 
show no effects for guns or either type of license. Coefficient estimates 
on total restricted firearms show null effects for first degree murders, 
negative effects on second degree murders (coefficient equals $-2.176$, 
significant at $5\%$ level), and positive effects on manslaughter 
(coefficient equals 0.964, significant at $10\%$ level). Our coefficient 
estimates on PALs suggest a one standard deviation increase would translate 
to $0.647$ (significant at $10\%$ level) more first degree murders, $-1.066$ 
(significant at $1\%$ level) second degree murders, and no effects on 
manslaughter. RPALs have null effects on first and second degree murders 
and negative effects on  manslaughters (coefficient equals $-0.660$, 
significant at $5\%$ level).

Panels 4 and 5 both include PALs and RPALs, while Panel 4 also includes 
total restricted firearms and Panel 5 additionally disaggregates restricted 
firearms into handguns, rifles, and other guns. Panel 5 is similar to the effects 
in panel 4 in many ways with respect to licenses variables 
which display both parameter stability (magnitude and sign). 
Allowing for heterogeneity in our firearms refines the 
estimates on licenses enough to show RPALs to be increasing 
using firearms (coefficient estimate equalling $1.917$ and 
significant at $10\%$ level) and PALs to be decreasing 
documentation issues (coefficient estimates equalling 
$-1.526$ significant at the $10\%$ level). Our coefficient estimate on RPALs is 
statistically significant and negatively related to manslaughter.

While panel four shows no effects across non-homicide crimes, panel 
five finds effects when breaking firearms out by their types. A one 
standard deviation increase in rifles corresponds to an increase in 
$22.352$ firearms violations (approximately four times the dependent 
variable mean).  This is statistically significant at the 1\% level.
Columns 2 through 5 show that the increase in firearms 
violations due to rifles comes from discharges (coefficient estimate 
equalling $10.859$),  using firearms (coefficient estimates $6.663$) 
and improper storage (coefficient estimates equalling $2.643$). 
Handguns seemingly offset using firearms and improper storage, 
with coefficient estimates being $-6.673$ and $-1.607$ in their 
respective models. Handguns do correspond to greater documentation 
issues however, with statistically significant and positive coefficient 
estimates equal to $4.563$.

Homicides generally are unaffected by any of our licenses or guns 
variables, with all coefficient estimates being statistically 
insignificant. First degree murders are positively related to handguns 
(coefficient equalling $1.658$) and PALs (coefficient equalling 0.601), 
both of which are statistically significant at the $10\%$ level. Second 
degree murders are decreasing in rifles (coefficient equalling $-2.257$, 
significant at the $5\%$ level) and PALs (coefficient equalling $-1.137$, 
significant at the $1\%$ level). Manslaughters are increasing in rifles 
(coefficient equalling $1.636$, significant at the $1\%$ level), other 
guns (coefficient equalling $0.325$, significant at the $1\%$ level) and 
PALs (coefficient equalling $0.328$, significant at the $1\%$ level), but 
decreasing in RPALs (coefficient equalling $-0.662$, significant at the 
$5\%$ level). It should be noted that within CMA correlation of rifles 
and handguns is 0.6519.

The coefficient on standardized rifles in the manslaughter specification is
notably larger than those observed in the neighbouring homicide regressions. 
One possible explanation is that manslaughter is a relatively rare outcome 
with substantial year-to-year variation across CMAs, which can amplify 
estimated associations when using standardized regressors. In addition, 
manslaughter incidents may be more sensitive to local fluctuations or a 
small number of high-incidence observations compared to more aggregated 
homicide measures. While the estimates are statistically precise, the 
magnitude should be interpreted with caution, as it may reflect the 
influence of a limited number of high-variation CMA-year observations 
rather than a broad underlying relationship.

This analysis suggests that the overall relationship between registered
and restricted firearms, or firearms licences, and deaths or crimes is 
fairly minimal. Moreover, we have not attempted to control for multiple
testing, or for the small number of clusters.  That said, we do find small but 
statistically significant and persistent relationships between PAL licences and 
total firearms deaths, suggesting about $3,846$ fewer license 
holders (per $100,000$) would reduce one firearms death (per $100,000$). However,
the sample mean is $4,639$ PAL holders per 100,000. We also note a persistent and 
relatively large relationship between other guns (neither rifles nor handguns) and 
adverse outcomes. The sample mean of this variable is just 15 per $100,000$ so 
these guns are quite rare in Canada.

\section{Conclusions}\label{section:concl}
%%%%%%%%%%%%%

This paper aims to document, for the first time, the landscape of firearms licenses
and registered and restricted firearms across Canada. In so doing, we make an empirical 
contribution to the debate on gun regulation in Canada.  This is particularly relevant 
given the recently enacted freeze on handguns and proposed assault rifle buybacks. This 
importantly offers a discussion on the role of regulating firearms in Canada, itself a 
highly dynamic and contentious subject, based on available data. More broadly, we 
contribute to a literature often hampered by a deficiency of accurate data or 
direct measures. 

Making use of a unique dataset on Canadians' restricted and registered firearms and 
those with valid licenses, we investigate the relationship between these firearms 
variables and deaths, crime and homicides. Importantly for us, this dataset contains
all people who can acquire a firearm through the legal market and the total counts 
of legal handguns. We provide descriptive evidence of firearms prevalence at the 
national level as well as in simple provincial regressions. Using a fixed effects 
model to control for heterogeneity across space in Canada, we estimate the impacts 
of licenses and restricted guns on firearms deaths, crimes, and homicides at the 
CMA/CA level and the CMA level. In general, licenses appear to be unrelated to 
firearms-related deaths. Other guns (commercial versions, some shotguns, air guns, 
submachine guns, machine guns, and some combination guns) appear to be increasing 
all firearms deaths as well as the subsets of firearms related deaths whose intent 
is thought to be assaults and self harm. Legal handguns are unrelated to total 
firearms deaths, deaths by assaults and self-harm. Legal handguns are, however, 
positively related with total handgun deaths. These relationships are small, with 
roughly a doubling of the stock of guns needed to increase deaths by 1 per 100,000.  

With respect to firearms-related crimes, we find actual counts of rifles to be 
increasing total firearms violations with licenses playing a much lesser role. While
total homicides appear to be unrelated to different types of firearms and different 
types of licenses, the different types of murder - first degree, second degree and 
manslaughter - have substantial heterogeneity. In aggregate, these results suggest 
that the Canadian style of gun control ``works''. When analyzing a period of nearly 
50\% growth in license holders, we fail to see systematic increases in gun deaths or
gun crimes. This suggests that the system of hands-on training, background checks, and 
safe storage enables private ownership without a negative externality.

The location of guns and their owners are unlikely to be randomly assigned across Canada. 
Future research should aim to incorporate both economic models and plausibly exogenous 
variation in the assignment of guns and licenses -- be they tightening or loosening of 
restrictions -- to yield credible estimates on the impacts of guns on outcomes like 
crime and suicides. Being able to better match the guns and licenses data with quality 
data on causes of deaths -- currently unavailable to us researchers -- represents a tangible 
improvement. Note, this is not the same as saying that licensing reduces 
crime. In other words, overall there does not appear to be a strong relationship between
crime and legal ownership. 

\pagebreak

\singlespacing

\phantomsection
\addcontentsline{toc}{section}{References}
\bibliographystyle{cje}
\bibliography{guns_bibliography}

%Giving the figures a bit more breathing room 
\newgeometry{tmargin=0.5in,bmargin=0.75in,lmargin=0.75in,rmargin=0.75in}

\newpage
\phantomsection
\stepcounter{section}

\section{Figures}

%%%%%%%%%
% FIGURE: TRENDS WITH GUNS, LICENSES, HOMICIDES AND SUICIDES 
%%%%%%%%%

\clearpage

%

%%%%%%%%%
% FIGURE: TRENDS WITH GUNS, LICENSES, HOMICIDES AND SUICIDES 
%%%%%%%%%

\clearpage

%
%

%%%%%%%%%
% FIGURE: TRENDS WITH GUNS, LICENSES, HOMICIDES AND SUICIDES 
%%%%%%%%%

\clearpage

\clearpage

%%%%%%%%%
% FIGURE: 
%%%%%%%%%

%%%%%%%%%
% FIGURE: SHAPE FILE MAP LICENSES AND RESTRICTED FIREARMS MAJOR CITIES
%%%%%%%%%

\clearpage

%%%%%%%%%
% FIGURE: handguns and rifles by census subdivisions
%%%%%%%%%

\clearpage

\clearpage

\newpage
\phantomsection
\stepcounter{section}

\section{Tables}

\begin{table}[H]
\caption{Broad Differences between Canada and US Gun Regulation}
\label{tab:can_us_diff}
\renewcommand{\arraystretch}{1.0}
\begin{tabular*}{\linewidth}{ @{\extracolsep{\fill}}l*{2}{c}}
   \hline
   \hline
   \\
   \textit{Policy Characteristics} & Canada & USA \\
   \hline 
   \\
   \textbf{More Restrictive Access} \\ 
   \hspace{0.5cm} Must be over 18 to \textit{own} firearms 		& \greencheck 	& \greencheck 	\\ 
   \hspace{0.5cm} Required Background Checks 					& \greencheck 	& \yellowcheck 	\\
   \hspace{0.5cm} Mandatory Training and Licensing 				& \greencheck 	& \yellowcheck	\\
   \hspace{0.5cm} Safe Storage Laws								&\greencheck	& \yellowcheck \\ 
   \hspace{0.5cm} Firearms Registration 					    & \greencheck	& \redx \\
   \hspace{0.5cm} Restricted Ownership of Handguns          	& \greencheck    & \redx \\
   \hspace{0.5cm} Prohibited Ownership of \textit{Most} Magazine-fed Semi-Automatic Rifles  & \greencheck & \redx       \\
   \hspace{0.5cm} Prohibition of Short-barrelled Handguns   & \greencheck    & \redx \\[0.5cm]
   \textbf{Less Restrictive Access} \\ 
   \hspace{0.5cm} Open-Carry or Concealed-Carry of Handguns 	 & \redx        & \yellowcheck     \\
   \hspace{0.5cm} Second Amendment							 & \redx 		& \greencheck
   \\
   \hline
   \hline
\end{tabular*}
\begin{scriptsize}
%What the table describes
Green check marks (\greencheck) mean ``yes'', 
yellow check marks mean ``sometimes'' (\yellowcheck) 
and red check marks (\redcheck) mean ``no.'' 
\end{scriptsize}
\end{table}

\begin{landscape}

\begin{table}[H]
\scriptsize
\caption{ Summary Statistics for Census Metropolitan Areas and Census Agglomerations }
\label{tab:sstats_cmaca}
\renewcommand{\arraystretch}{1.0}
\begin{tabular*}{\hsize}{ @{\extracolsep{\fill}}l*{7}{c}}
\hline\hline
\\
\textit{Summary Statistics for CMA and CA}
                                                               & Mean & Std. Dev. &  Min & Median & Max & Observations\\
\hline
\\

\hspace{0.1cm}\textit{Main Independent Variables}                &      &      &      &     &  &     \\

\hspace{0.5cm}PAL per 100,000                                    & 8,241 & 4,138 & 790 & 7,307 & 24,627 & 1,089\\

\hspace{0.5cm}RPAL per 100,000                                   & 1,968 & 1,356 & 0 & 1,545 & 7,711 & 1,089\\

\hspace{0.5cm}Total Restricted Guns per 100,000                  & 3,658 & 3,563 & 228.5 & 2,929 & 77,084 & 1,089\\

\\
\hspace{0.1cm}\textit{Different Gun Types}                       &      &      &      &     &  &     \\

\hspace{0.5cm}Handguns per 100,000                               & 3,241 & 2,283 & 215.6 & 2,673 & 26,001 & 1,089\\

\hspace{0.5cm}Rifles per 100,000                                 & 403.5 & 2442.7  & 10.2  & 210.6  & 73939 & 1089\\

\hspace{0.5cm}Restricted Commercial Versions per 100,000         & 9.3 & 10.9 &  0 & 6 & 77.9 & 1089\\

\hspace{0.5cm}Shotguns per 100,000                               & 3.6 & 12.1 & 0 & 0 & 137.1 & 1,089\\

\hspace{0.5cm}Machine Guns per 100,000                           & 0.72 & 7.87 &  0 & 0 & 112.5 & 1089\\

\hspace{0.5cm}Submachine Guns per 100,000                        & 0.16 & 0.78 &  0 & 0  &  8.7 & 1089

\\
\hspace{0.1cm}\textit{Control Variables}                         &      &      &      &      & &    \\

\hspace{0.5cm}\% of Females                                      & 51.3 & 1.07 & 47 & 51 & 54 & 1089\\

\hspace{0.5cm}Avg. Age of Persons                                & 41 & 3.88 & 31 & 41 & 55 & 1,089\\

\hspace{0.5cm}\% of Married Persons                              & 37.7 & 6.5 & 18 & 39 & 60 & 1089\\

\hspace{0.5cm}\# of Persons Reporting EI                         & 124,997 & 367,064 & 860 & 23,000 & 3,658,630 & 1,089\\

\hspace{0.5cm}\# of Persons Reporting Federal Child Benefits     & 24,540 & 73,437 & 190 & 4,360 & 801,050 & 1,089\\
\\
\hline \hline

\end{tabular*}

\begin{footnotesize}
  Notes: Authors' calculations. %
  Years range from 2013 - 2019. %
  Guns data from \textit{Access to Information and Privacy} Request from the %
  RCMP; Crime and taxfiler information data from %
  \citet{StatCan_PopChange} and \citet{StatCan_TaxFilers}, respectively. %  
  Each observation is year by Census Metropolitan Area or Census Agglomeration. %
  EI stands for employment insurance. %
  PAL is an acronym for Possession and Acquisition License. %
  RPAL is an acronym for \textit{Restricted} PAL. %
  All tables constructed using \citet{Jann2005,Jann2007}.
\end{footnotesize}
\end{table}

\clearpage 

%
%

%%%%%%%%
% TABLE: Summary Statistics for CMAs
%%%%%%%%

\begin{table}[H]
\scriptsize
\caption{Summary Statistics for Census Metropolitan Areas}
\label{tab:sstats_cma}
\renewcommand{\arraystretch}{1.0}
\begin{tabular*}{\hsize}{@{\extracolsep{\fill}}l*{6}{c}}
\hline\hline
\\[-0.5em]
\textit{Summary Statistics for CMA}
    & Mean & Std. Dev. & Min & Median & Max & Observations \\
\hline
\\[-0.5em]
\hspace{0.1cm}\textit{Outcome Variables}
    & & & & & & \\
\hspace{0.5cm}Total Firearms Violations (Use, discharge, pointing) per 100,000
    & 5.65 & 4.77 & 0.0 & 4.3 & 27.9 & 225 \\
\hspace{0.5cm}Discharge Firearm with Intent per 100,000
    & 1.80 & 2.40 & 0.0 & 0.9 & 13.4 & 225 \\
\hspace{0.5cm}Using Firearm in Commission of Offence per 100,000
    & 1.55 & 1.61 & 0.0 & 1.0 & 9.1 & 225 \\
\hspace{0.5cm}Firearms Documentation Issues per 100,000
    & 1.37 & 2.98 & 0.0 & 0.2 & 19.5 & 225 \\
\hspace{0.5cm}Unsafe Firearms Storage per 100,000
    & 1.70 & 1.60 & 0.0 & 1.3 & 9.2 & 225 \\
\hspace{0.5cm}Homicides per 100,000
    & 1.62 & 1.29 & 0.0 & 1.4 & 9.0 & 225 \\
\hspace{0.5cm}1st Degree Murder per 100,000
    & 0.68 & 0.62 & 0.0 & 0.6 & 4.1 & 225 \\
\hspace{0.5cm}2nd Degree Murder per 100,000
    & 0.75 & 1.01 & 0.0 & 0.5 & 9.0 & 225 \\
\hspace{0.5cm}Manslaughter per 100,000
    & 0.19 & 0.30 & 0.0 & 0.0 & 1.6 & 225 \\
\\[-0.5em]
\hspace{0.1cm}\textit{Main Independent Variables}
    & & & & & & \\
\hspace{0.5cm}PAL per 100,000
    & 4,640 & 2,253 & 790 & 4,169 & 12,640 & 225 \\
\hspace{0.5cm}RPAL per 100,000
    & 1,339 & 662 & 115 & 1,217 & 3,383 & 225 \\
\hspace{0.5cm}Total Restricted Guns per 100,000
    & 2,526 & 1,059 & 229 & 2,449 & 7,111 & 225 \\
\\[-0.5em]
\hspace{0.1cm}\textit{Different Gun Types}
    & & & & & & \\
\hspace{0.5cm}Handguns per 100,000
    & 2,230 & 821 & 216 & 2,186 & 4,216 & 225 \\
\hspace{0.5cm}Rifles per 100,000
    & 280 & 338 & 12 & 220 & 2,849 & 225 \\
\hspace{0.5cm}Shotguns per 100,000
    & 4.45 & 9.29 & 0.0 & 1.3 & 71.0 & 225 \\
\hspace{0.5cm}Restricted Commercial Versions per 100,000
    & 8.68 & 6.56 & 0.0 & 6.6 & 34.7 & 225 \\
\hspace{0.5cm}Machine Guns per 100,000
    & 3.24 & 17.06 & 0.0 & 0.0 & 112.5 & 225 \\
\hspace{0.5cm}Submachine Guns per 100,000
    & 0.22 & 0.68 & 0.0 & 0.0 & 8.7 & 225 \\
\\[-0.5em]
\hspace{0.1cm}\textit{Control Variables}
    & & & & & & \\
\hspace{0.5cm}\% of Females
    & 51.57 & 0.55 & 50.0 & 52.0 & 53.0 & 225 \\
\hspace{0.5cm}Avg. Age of Persons
    & 40.25 & 1.99 & 36.0 & 40.0 & 44.0 & 225 \\
\hspace{0.5cm}\% of Married Persons
    & 38.55 & 4.88 & 24.0 & 40.0 & 44.0 & 225 \\
\hspace{0.5cm}\# of Persons Reporting EI
    & 424,345 & 699,476 & 55,060 & 140,770 & 3,658,630 & 225 \\
\hspace{0.5cm}\# of Persons Reporting Federal Child Benefits
    & 83,071 & 140,892 & 10,760 & 30,540 & 801,050 & 225 \\
\\[-0.5em]
\hline\hline
\end{tabular*}

\begin{footnotesize}
  Notes: Authors' calculations. %
  Years range from 2013 - 2019. %
  Guns data from \textit{Access to Information and Privacy} Request from the %
  RCMP; Crime and taxfiler information data from %
  \citet{StatCan_Crime} and \citet{StatCan_TaxFilers}, respectively. %  
  Each observation is year by Census Metropolitan Area. %
  EI stands for employment insurance. %
  PAL is an acronym for Possession and Acquisition License. %
  RPAL is an acronym for \textit{Restricted} PAL. %
\end{footnotesize}
\end{table}

\end{landscape}

\clearpage

\begin{landscape}

%\subsection*{Provincial Tables}

%%%%%%%%
% TABLE: 
%%%%%%%%

\begin{table}[H]
\caption{ Provincial Firearms Crimes, Homicides, Firearms Licenses and Registered Restricted Firearms }
\label{tab:prov_crimes}
\scriptsize
\renewcommand{\arraystretch}{1.0}
\begin{tabular*}{\linewidth}{ @{\extracolsep{\fill}}l*{9}{c}}
\hline\hline
\\

 &\multicolumn{9}{c}{ \textit{Dependent Variable}: \textsc{Total Crimes (per 100,000)} } \\
\cline{2-9}
\\

                      & Total Firearms & Firearms & Using & Documentation & Improper & 1st Degree & 2nd Degree &  \\ 
\textit{Model/Independent Variables} & Violations & Discharge & Firearms & Issue & Storage & Murder & Murder & Manslaughter \\ 
\cline{2-9}
\\

\multicolumn{4}{l}{\textsc{Total Licenses (per 100,000)}} &&&& \\[0.05em]
\\
\hfill $\hat{\beta}_{1} \times 1000$            & 0.128*** & 0.040*** & 0.024*** & -2.601*** & 0.064*** & 0.003    & 0.012*** & 0.003***\\
                                         & (0.000)    & (0.000)    & (0.000)    & (0.000)    & (0.000)    & (0.000)    & (0.000)    & (0.000)   \\[0.5em]

\hfill  $\hat{\beta}_{2}$ (annual $\% \Delta$)  & -0.031    & -0.017    & -0.006    & -0.116    & 0.004    & -0.002    & -0.001    & 0.001   \\
                                         & (0.028)    & (0.018)    & (0.005)    & (1.030)    & (0.016)    & (0.002)    & (0.004)    & (0.001)   \\
\\
Independent Variable Mean                & 8620.9    & 8620.9    & 8620.9    & 8620.9    & 8620.9    & 8620.9    & 8620.9    & 8620.9   \\
\hdashline
\\

\multicolumn{4}{l}{\textsc{Total Registered Restricted Firearms (per 100,000)}} &&&& \\[0.05em]
\\
\hfill $\hat{\beta}_{1} \times 1000$            & 0.003*** & 0.001*** & 0.000*** & -0.005    & 0.001*** & 0.000*** & 0.000*** & 0.000***\\
                                         & (0.000)    & (0.000)    & (0.000)    & (0.010)    & (0.000)    & (0.000)    & (0.000)    & (0.000)   \\[0.5em]

\hfill  $\hat{\beta}_{2}$ (annual $\% \Delta$)  & -0.139    & -0.015    & -0.016    & 8.068    & -0.129    & 0.003    & -0.001    & 0.002   \\
                                         & (0.116)    & (0.048)    & (0.028)    & (8.548)    & (0.147)    & (0.013)    & (0.017)    & (0.008)   \\
\\
Independent Variable Mean                & 2775.8    & 2775.8    & 2775.8    & 2775.8    & 2775.8    & 2775.8    & 2775.8    & 2775.8   \\
Dependent Variable Mean                  & 9.4    & 3.5    & 1.7    & 30.5    & 3.8    & .7    &  .9    &  .3   \\
Observations                             &  60    &  60    &  60    &  60    &  60    &  60    &  60    &  60   \\
\hline \hline
\end{tabular*}

 \begin{footnotesize}
  Notes: Authors' calculations. %
  Guns data from \textit{Access to Information and Privacy} Request from the %
  RCMP; Deaths data from \citet{StatCan_CauseDeath}. %
  Observations are Province-by-year between 2014 and 2019. There %
  are ten provinces and six years. %
  Models are: $y_{p,y} = \beta_{0} + \beta_{1} x_{p,y} + \beta_{2} ( \Delta_{\%} x_{p,y}) + \varepsilon_{p,y}$. %
  All models are estimated use ordinary least squares and %
  standard errors clustered at the provincial level. %
  Provincial population are used as weights. %
  Columns vary by the dependent variable used in the model. %
  Columns 1 is Criminal Code violation 150 (per 100,000) - total firearms violations, use of, discharge, pointing. %
  Column 2 is Criminal Code violation 1450 (per 100,000) - discharge firearms with intent. %
  Column 3 is Criminal Code violation 1455 (per 100,000) - using firearms in commission of offence. %
  Column 4 is Criminal Code violation 3390 (per 100,000) - firearms documentation or administrations. %
  Column 5 is Criminal Code violation 3395 (per 100,000) - unsafe storage of firearms. %  
  Columns 6 is Criminal Code violation 1110 (per 100,000) - Murder, first degree. %
  Columns 7 is Criminal Code violation 1120 (per 100,000) - Murder, second degree. %
  Columns 8 is Criminal Code violation 1130 (per 100,000) - Manslaughter. % 
  Each panel represents a different model, each with two estimated %
  coefficients. Panel 1 regresses outcome variables on Total Licenses; %
  Panel 2 regresses outcome variables on Total Restricted Guns. %
  $\hat{\beta}_{1}$ is multiplied by 1000 for interpretation. %
  * significant at 10\%, ** significant at 5\%, *** significant at 1\%. 
\end{footnotesize}
%\end{sidewaystable}
\end{table}

\end{landscape}

\clearpage 

%
%

%%%%%%%%
% TABLE: 
%%%%%%%%

\begin{landscape}

\begin{table}[H]
\caption{ Provincial Homicides, Suicides, Firearms Licenses and Registered Restricted Firearms }
\label{tab:prov_ols}
\scriptsize
\renewcommand{\arraystretch}{1.0}
\begin{tabular*}{\linewidth}{ @{\extracolsep{\fill}}l*{13}{c}}
\hline\hline
\\

 &\multicolumn{13}{c}{ \textit{Dependent Variable}: \textsc{Total Deaths (per 100,000) by Sample and Type} } \\
\cline{2-13}
\\

& \multicolumn{4}{c}{ \textsc{Both Sexes} } & \multicolumn{4}{c}{ \textsc{Males} } & \multicolumn{4}{c}{ \textsc{Females} } \\ 
\cline{2-5} \cline{6-9} \cline{10-13}
\\

\textit{Model/Independent Variables} & Suicides & Homicides & Accidents & Other & Suicides & Homicides & Accidents & Other & Suicides & Homicides & Accidents & Other \\ 
\cline{2-13}
\\

\multicolumn{4}{l}{\textsc{Total Licenses (per 100,000)}} &&&&&&&&&& \\[0.05em]
\\
\hfill $\hat{\beta}_{1} \times 1000$            & 0.057*** & 0.003    & 0.025    & 0.342**  & 0.047*** & 0.002    & 0.051**  & 0.217*** & 0.010*** & 0.001    & -0.026*   & 0.129   \\
                                         & (0.000)    & (0.000)    & (0.000)    & (0.000)    & (0.000)    & (0.000)    & (0.000)    & (0.000)    & (0.000)    & (0.000)    & (0.000)    & (0.000)   \\[0.5em]

\hfill  $\hat{\beta}_{2}$ (annual $\% \Delta$)  & 0.039    & 0.005    & -0.041    & -0.388**  & 0.057    & 0.004    & -0.047    & -0.428**  & 0.023*   & 0.006    & -0.034    & -0.349   \\
                                         & (0.024)    & (0.006)    & (0.062)    & (0.193)    & (0.037)    & (0.009)    & (0.079)    & (0.170)    & (0.013)    & (0.004)    & (0.064)    & (0.222)   \\
\\
Independent Variable Mean                & 8620.9    & 8620.9    & 8620.9    & 8620.9    & 17354.4    & 17354.4    & 17354.4    & 17354.4    & 17134.2    & 17134.2    & 17134.2    & 17134.2   \\
\hdashline
\\

\multicolumn{4}{l}{\textsc{Total Registered Restricted Firearms (per 100,000)}} &&&&&&&&&& \\[0.05em]
\\
\hfill $\hat{\beta}_{1} \times 1000$            & 0.000    & 0.000**  & 0.002**  & 0.002    & 0.000    & 0.000**  & 0.002*** & 0.003*   & 0.000    & 0.000**  & -0.001    & -0.001   \\
                                         & (0.000)    & (0.000)    & (0.001)    & (0.003)    & (0.000)    & (0.000)    & (0.000)    & (0.002)    & (0.000)    & (0.000)    & (0.000)    & (0.002)   \\[0.5em]

\hfill  $\hat{\beta}_{2}$ (annual $\% \Delta$)  & -0.104    & 0.029    & 0.017    & -0.340    & -0.154    & 0.035    & -0.035    & -0.577    & -0.041    & 0.021    & 0.083    & -0.106   \\
                                         & (0.121)    & (0.032)    & (0.210)    & (1.117)    & (0.201)    & (0.050)    & (0.214)    & (1.142)    & (0.051)    & (0.015)    & (0.235)    & (1.140)   \\
\\
Independent Variable Mean                & 2775.8    & 2775.8    & 2775.8    & 2775.8    & 5580.6    & 5580.6    & 5580.6    & 5580.6    & 5524.4    & 5524.4   & 5524.4    & 5524.4  \\
Dependent Variable Mean                  & 12.4    & 1.4    &  37    & 165.8    & 18.9    &   2    & 43.1    & 147.9    & 5.9    & .7   &  31    & 183.5   \\
Observations                             &  60    &  60    &  60    &  60    &  60    &  60    &  60    &  60    &  60    &  60    &  60    &  60   \\
\hline \hline
\end{tabular*}

 \begin{footnotesize}
  Notes: Authors' calculations. %
  Guns data from \textit{Access to Information and Privacy} Request from the %
  RCMP; Deaths data from \citet{StatCan_CauseDeath}. %
  Observations are Province-by-year between 2014 and 2019. There %
  are ten provinces and six years. %
  Models are: $y_{p,y} = \beta_{0} + \beta_{1} x_{p,y} + \beta_{2} ( \Delta_{\%} x_{p,y}) + \varepsilon_{p,y}$. %
  All models are estimated use ordinary least squares and %
  standard errors clustered at the provincial level. %
  Provincial population (by sample) are used as weights. %
  Each panel represents a different model, each with two estimated %
  coefficients. Panel 1 regresses outcome variables on Total Licenses; %
  Panel 2 regresses outcome variables on Total Restricted Guns. %
  Columns 1 to 4, 5 to 8, 9 to 12, are over the samples of both sexes, %
  only males, and only females, respectively. %
  $\hat{\beta}_{1}$ is multiplied by 1000 for interpretation. %
  Columns vary the sample and the deaths by category. %
  Columns 1, 5, and 9 are the total suicides per 100,000. %
  Columns 2, 6, and 10 are the total homicides per 100,000. %
  Columns 3, 7, and 11 are the total accidental deaths per 100,000. %
  Columns 4, 8, and 12 are the total other deaths per 100,000. %  
  * significant at 10\%, ** significant at 5\%, *** significant at 1\%. 
\end{footnotesize}
%\end{sidewaystable}
\end{table}

\end{landscape}

\clearpage

\begin{table}[H]
 \caption{Firearms Deaths}
 \footnotesize
 \label{tab:rdc_all_deaths}
 \renewcommand{\arraystretch}{.75}
\begin{tabular*}{\linewidth}{ @{\extracolsep{\fill}}l*{6}{c}}
\hline\hline
\\

\textit{Independent Variables} \\ 
\textit{(per 100,000}) &\multicolumn{5}{c}{ \textit{Dependent Variable}: \textsc{Total Firearms Deaths (per 100,000)} } \\
\cline{2-7}
\\

\textsc{Panel 1} && \\
Total Restricted Guns        & 0.41    & 0.26    & 0.23    & 0.20    & -0.01   \\
                            & (0.23)  & (0.14)  & (0.16)  & (0.15)  & (0.03)   \\

\hdashline
\\

\textsc{Panel 2} && \\
All Licenses                 & 0.29*** & 0.26*** & 0.29*** & 0.30*** & 0.06   \\
                            & (0.04)  & (0.04)  & (0.04)  & (0.05)  & (0.06)   \\

\hdashline
\\

\textsc{Panel 3} && \\
Total Restricted Guns        & 0.00    & -0.00   & 0.04    & 0.04    & -0.02   \\
                            & (0.04)  & (0.03)  & (0.04)  & (0.04)  & (0.02)   \\[0.5em]

All Licenses                 & 0.29*** & 0.27*** & 0.28*** & 0.29*** & 0.06   \\
                            & (0.03)  & (0.04)  & (0.03)  & (0.04)  & (0.06)   \\
\hdashline
\\

\textsc{Panel 4} && \\
Total Restricted Guns        & -0.01   & 0.01    & 0.03    & 0.02    & -0.02   \\
                            & (0.01)  & (0.02)  & (0.02)  & (0.01)  & (0.02)   \\[0.5em]

PAL                         & 0.27*** & 0.28*** & 0.27*** & 0.25*** & 0.08   \\
                            & (0.03)  & (0.05)  & (0.04)  & (0.04)  & (0.11)   \\[0.5em]

RPAL                        & 0.46*   & 0.18    & 0.36*   & 0.53**  & -0.02   \\
                            & (0.21)  & (0.11)  & (0.16)  & (0.19)  & (0.49)   \\
\hdashline
\\

\textsc{Panel 5} && \\
Handguns                    & 0.09    & 0.11    & 0.17    & 0.12**  & -0.02   \\
                            & (0.08)  & (0.10)  & (0.10)  & (0.04)  & (0.06)   \\[0.5em]

Rifles                      & -0.06   & -0.04   & -0.03   & -0.03   & -0.02   \\
                            & (0.04)  & (0.04)  & (0.03)  & (0.02)  & (0.02)   \\[0.5em]

Other guns                  & 0.71    & 1.14    & 1.22    & 0.92    & 1.97*** \\
                            & (2.00)  & (1.68)  & (1.72)  & (1.84)  & (0.48)   \\[0.5em]

PAL                         & 0.26*** & 0.27*** & 0.26*** & 0.25*** & 0.08   \\
                            & (0.02)  & (0.04)  & (0.04)  & (0.04)  & (0.12)   \\[0.5em]

RPAL                        & 0.34    & 0.07    & 0.20    & 0.40*   & -0.01   \\
                            & (0.21)  & (0.14)  & (0.19)  & (0.19)  & (0.53)   \\[0.5em]

P-value (All Guns equal to 0) & 0.718   & 0.487   & 0.454   & 0.596   & 0.002   \\                 
\\                        
\hdashline \\

Province Fixed Effects (FE)             &  Yes    &   No    &  Yes    &  Yes    &  Yes   \\
Demographic Controls                    &   No    &  Yes    &  Yes    &  Yes    &  Yes   \\
Year FE                                 &   No    &   No    &   No    &  Yes    &  Yes   \\
CMA and CA FE                           &   No    &   No    &   No    &   No    &  Yes   \\\\[0.5em]
Observations                            & 1015    & 1015    & 1015    & 1015    & 1015   \\
\\
\hline \hline
\end{tabular*}
 \begin{scriptsize}
 \begin{singlespace}
  Notes: All coefficients reported have been multiplied by 1000. Authors' calculations. %
  Deaths data from the Vital Statistics Death Database (VSDD) accessed in Statistics Canada's \textit{Research Data Center}. %
  Years range from 2013 - 2019. %
  Each observation is year by Census Metropolitan Area (CMA) \textit{or} Census Agglomeration (CA). %
  All equations are estimated using ordinary least squares with weights applied. %
  Panels vary the different firearms variables used in each model.
  Columns vary by their inclusion of various control variables %
  and fixed effects (FE). %
  Columns (2) to (5) control for demographic characteristics at %
  CMA or CA level. %
  These variables include: the average age of a person, the percentage of %
  people that are female, the percentage of people that are married, the %
  number of people receiving (un)Employment Insurance (EI) Benefits, and the number of people %
  receiving federal child benefits. %
  PAL is an acronym for Possession and Acquisition License; RPAL is a Restricted PAL. 
  Standard errors are constructed using a CRVE and clustered %
  at the province-level. %
  * significant at 10\%, ** significant at 5\%, *** significant at 1\%. 
  \end{singlespace}
\end{scriptsize}
\end{table}

\clearpage

\clearpage

\begin{table}[H]
 \caption{ Firearms Deaths by Intent of Death}
 \footnotesize
\renewcommand{\arraystretch}{.75}
 \label{tab:rdc_deaths_intent}
\begin{tabular*}{\linewidth}{ @{\extracolsep{\fill}}l*{3}{c}}
\hline\hline
\\
\textit{Independent Variables} &\multicolumn{3}{c}{\textit{Dependent Variables}: \textsc{Deaths by Intent} {\scriptsize \textit{(per 100,000)}} }  \\[0.5em]
\textit{(per 100,000)}	 & Assaults & Self-Harm & Unknown \\
\cline{2-4} \\

\textsc{Panel 1} && \\
Total Restricted Guns per 100,000        & -0.59    & -19.08    & 0.74  \\
                                         & (9.36)   & (11.04)   & (1.16) \\[0.5em]

All Licenses per 100,000                 & 7.10     & 56.05     & -3.54 \\
                                         & (19.40)  & (64.53)   & (6.47)  \\
\\
P-value (Licenses + Guns = 0)            & .803    & .603    & .638    \\
\\
\hdashline
\\

\textsc{Panel 2} && \\
Total Restricted Guns        & 3.84      & -22.86**    & 1.45      \\
                             & (10.27)   & (9.80)      & (1.58)    \\[0.5em]

PAL                          & 32.05     & 34.70       & 0.47     \\
                             & (24.72)   & (106.91)    & (8.72)   \\[0.5em]

RPAL                         & -144.18   & 185.46      & -27.80   \\
                             & (215.19)  & (328.43)    & (36.73)  \\
\\
P-value (RPAL + PAL + Guns = 0) & .596  & .453        & .436  \\
\\
\hdashline
\\

\textsc{Panel 3} && \\
Handguns                     & 25.88     & -46.42      & 4.98      \\
                             & (28.33)   & (39.77)     & (3.78)    \\[0.5em]

Rifles                       & 1.07      & -21.68*     & 1.17     \\
                             & (6.01)    & (10.75)     & (1.14)   \\[0.5em]

Other guns  				 & 724.14**  & 1273.45***  & 4.27    \\
			                 & (247.66)  & (287.61)    & (25.17)  \\[0.5em]

PAL                          & 32.48     & 32.29       & 0.64    \\
                             & (26.04)   & (107.65)    & (8.74)   \\[0.5em]

RPAL                         & -155.95   & 212.23      & -30.48   \\
                             & (226.81)  & (361.63)    & (37.03)   \\[0.5em]
                             
P-value (All Guns Equal 0)              & .013    & .001    & .724     \\[0.5em]

Province, Year and CMA FE   &  Yes    &  Yes    &  Yes    \\
Demographic Controls        &  Yes    &  Yes    &  Yes    \\
Observations                & 1015    & 1015    & 1015   \\
\\
\hline \hline
\end{tabular*}

 \begin{scriptsize}
 \begin{singlespace}

  Notes: All coefficients reported have been multiplied by 1000. Authors' calculations. %
  Deaths data from the Vital Statistics Death Database (VSDD) accessed in Statistics Canada's \textit{Research Data Center}. %
  Years range from 2013 - 2019. %
  Each observation is year by Census Metropolitan Area (CMA) \textit{or} Census Agglomeration (CA). %
  All equations are estimated using ordinary least squares with weights applied. %
  Panels vary the different firearms variables used in each model.
  Columns vary by their inclusion of various control variables %
  and fixed effects (FE). %
  Columns (2) to (5) control for demographic characteristics at %
  CMA or CA level. %
  These variables include: the average age of a person, the percentage of %
  people that are female, the percentage of people that are married, the %
  number of people receiving (un)Employment Insurance (EI) Benefits, and the number of people %
  receiving federal child benefits. %
  PAL is an acronym for Possession and Acquisition License; RPAL is a Restricted PAL. 
  Standard errors are constructed using a CRVE and clustered %
  at the province-level. %
  * significant at 10\%, ** significant at 5\%, *** significant at 1\%. 
  
 \end{singlespace}
\end{scriptsize}
\end{table}

\begin{table}[H]
 \caption{ Firearms Deaths by Different Types of Firearms}
 \footnotesize
 \label{tab:rdc_deaths_types}
 \renewcommand{\arraystretch}{.75}
\begin{tabular*}{\linewidth}{ @{\extracolsep{\fill}}l*{4}{c}}
\hline\hline
\\
\textit{Independent Variables} &\multicolumn{4}{c}{ \textit{Dependent Variables}: \textsc{Types of Firearms Deaths} \textit{(per 100,000)}} \\[0.5em]
\textit{(per 100,000)} & Handguns & Rifles & Unspecified & Uncategorised  \\
\cline{2-5} \\

\textsc{Panel 1} && \\
Total Restricted Guns per 100,000        & 5.27    & -7.56    & -18.33    & -1.70   \\
                                         & (4.17)  & (33.81)  & (23.26)   & (1.93)   \\[0.5em]

All Licenses per 100,000                 & 10.55   & -21.57   & 75.69     & 5.06   \\
                                         & (20.99) & (38.20)  & (42.28)   & (6.46)   \\
\\
P-value (Licenses + Guns = 0)            & .542    & .514     & .202      & .640   \\
\hdashline
\\

\textsc{Panel 2} && \\
Total Restricted Guns                    & 5.58**   & -25.89*** & 2.34     & -0.39   \\
                                         & (2.31)   & (7.25)    & (12.62)  & (1.37)   \\[0.5em]

PAL                                      & 12.31    & -124.95   & 192.30*  & 12.45   \\
                                         & (14.85)  & (134.02)  & (91.65)  & (9.62)   \\[0.5em]

RPAL                                     & -0.16    & 605.28    & -631.43  & -39.80   \\
                                         & (111.82) & (699.15)  & (423.27) & (34.64)   \\
\\
P-value (RPAL + PAL + Guns = 0)          & .881     & .455      & .269     & .355   \\
\\
\hdashline
\\

\textsc{Panel 3} && \\
Handguns                                 & 17.97*   & -57.61    & 21.75    & -2.32   \\
                                         & (8.01)   & (42.38)   & (38.36)  & (7.78)   \\[0.5em]

Rifles                                   & 4.50**   & -22.88*** & -1.28    & -0.22   \\
                                         & (1.55)   & (6.81)    & (9.83)   & (1.01)   \\[0.5em]

Other guns                               & 41.31    & -537.96*** & 2467.60*** & -30.91   \\
                                         & (88.00)  & (113.74)   & (391.03)   & (40.28)   \\[0.5em]

PAL                                      & 12.90    & -126.02    & 190.92*   & 12.39   \\
                                         & (14.51)  & (133.24)   & (88.87)   & (9.86)   \\[0.5em]

RPAL                                     & -9.34    & 625.69     & -629.08   & -38.53   \\
                                         & (108.25) & (684.60)   & (388.78)  & (36.86)   \\[0.5em]

P-value (All Guns Sum to 0)              & .511     & .001       & .000      & .411   \\[0.5em]

Province, Year and CMA FE                & Yes      & Yes        & Yes       & Yes   \\
Demographic Controls                     & Yes      & Yes        & Yes       & Yes   \\[0.5em]
Observations                             & 1015     & 1015       & 1015      & 1015   \\
\\
\hline \hline
\end{tabular*}

 \begin{scriptsize}
 \begin{singlespace}
  Notes: Authors' calculations. %
  Deaths data from the Vital Statistics Death Database (VSDD) accessed in Statistics Canada's \textit{Research Data Center}. %
  Years range from 2013 - 2019. %
  Each observation is year by Census Metropolitan Area (CMA) \textit{or} Census Agglomeration (CA). %
  All equations are estimated using ordinary least squares with weights applied. %
  Panels vary the different firearms variables used in each model.
  Columns vary by their inclusion of various control variables %
  and fixed effects (FE). %
  Columns (2) to (5) control for demographic characteristics at %
  CMA or CA level. %
  These variables include: the average age of a person, the percentage of %
  people that are female, the percentage of people that are married, the %
  number of people receiving (un)Employment Insurance (EI) Benefits, and the number of people %
  receiving federal child benefits. %
  PAL is an acronym for Possession and Acquisition License; RPAL is a Restricted PAL. 
  Standard errors are constructed using a CRVE and clustered %
  at the province-level. %
  * significant at 10\%, ** significant at 5\%, *** significant at 1\%. 
 \end{singlespace}
 \end{scriptsize}
\end{table}

\clearpage 

%%

%
%

%%%%%%%%
% TABLE: 
%%%%%%%%

\begin{landscape}  
  
\begin{table}[H]
\caption{ Crimes, Licenses, Restricted Registered Guns in CMAs}
\label{tab:cma_firecrimes} 
\tiny
\renewcommand{\arraystretch}{.75}
\begin{tabular*}{\linewidth}{ @{\extracolsep{\fill}}l*{10}{c}}
\hline\hline
\\
&\multicolumn{10}{c}{ \textit{Dependent Variables}: \textsc{Total Offences (per 100,000) by Type} } \\
\cline{2-10}
\\

                      & Total Firearms & Firearms & Using & Documentation & Improper & & 1st degree & 2nd degree &  \\ 
\textit{Model/Independent Variables} & Violations & Discharge & Firearms & Issue & Storage & Homicides & Murder & Murder & Manslaughter \\ 
\cline{2-10}

\multicolumn{4}{l}{\textsc{Model 1}} &&&&& \\
Total Restricted Guns per 100,000 (Standardized) & 15.785    & 7.595    & 2.304    & 3.341    & 1.924*   & 0.326    & 1.480    & -1.632**  & 0.483   \\
                                         & (12.153)    & (5.868)    & (4.320)    & (2.101)    & (1.008)    & (1.389)    & (0.865)    & (0.679)    & (0.456)   \\
\\
\hdashline
\\

\multicolumn{4}{l}{\textsc{Model 2}} &&&&& \\
All Licenses per 100,000 (Standardized)  & -0.300    & -0.344    & 0.376    & -0.993    & 0.450    & -0.265    & 0.699*** & -1.051*** & 0.091   \\
                                         & (2.213)    & (1.261)    & (0.647)    & (0.639)    & (0.406)    & (0.235)    & (0.189)    & (0.220)    & (0.140)   \\
\\
\hdashline
\\

\multicolumn{4}{l}{\textsc{Model 3}} &&&&& \\
Total Restricted Guns per 100,000 (Standardized) & 17.066    & 8.315    & 2.276    & 4.090    & 1.830    & 0.485    & 1.226    & -1.208    & 0.470   \\
                                         & (12.329)    & (6.058)    & (4.487)    & (2.270)    & (1.124)    & (1.401)    & (0.797)    & (0.881)    & (0.437)   \\
[1em]
All Licenses per 100,000 (Standardized)  & -2.673    & -1.500    & 0.059    & -1.561**  & 0.196    & -0.332*** & 0.529*   & -0.883**  & 0.026   \\
                                         & (1.790)    & (1.214)    & (0.651)    & (0.633)    & (0.349)    & (0.081)    & (0.237)    & (0.299)    & (0.144)   \\
\\
\hdashline
\\

\multicolumn{4}{l}{\textsc{Model 4}} &&&&& \\
Total Restricted Guns per 100,000 (Standardized) & 16.981    & 9.112    & 1.079    & 3.636    & 1.183    & 0.615    & 1.831    & -2.176**  & 0.964*  \\
                                         & (13.143)    & (5.243)    & (4.735)    & (2.678)    & (0.897)    & (1.704)    & (1.124)    & (0.831)    & (0.515)   \\
[1em]
PAL per 100,000 (Standardized)           & -2.246    & -0.969    & -0.363    & -1.451    & -0.061    & -0.231    & 0.647*   & -1.066*** & 0.191   \\
                                         & (1.569)    & (1.178)    & (0.634)    & (0.800)    & (0.426)    & (0.209)    & (0.281)    & (0.292)    & (0.132)   \\
[1em]
RPAL per 100,000 (Standardized)          & -0.605    & -1.481    & 1.632    & 0.193    & 0.926    & -0.266    & -0.675    & 1.070    & -0.660** \\
                                         & (4.711)    & (2.150)    & (1.574)    & (1.275)    & (0.625)    & (0.686)    & (0.569)    & (0.615)    & (0.276)   \\
\\
\hdashline
\\

\multicolumn{4}{l}{\textsc{Model 5}} &&&&& \\
Handguns per 100,000 (Standardized)      & -4.486    & -1.064    & -6.673**  & 4.563**  & -1.607*   & 1.117    & 1.658*   & -0.157    & -0.376   \\
                                         & (5.028)    & (3.658)    & (2.868)    & (1.880)    & (0.775)    & (1.438)    & (0.794)    & (0.687)    & (0.496)   \\[0.5em]

Rifles per 100,000 (Standardized)        & 22.352*** & 10.859*** & 6.663*   & 1.321    & 2.643*** & 0.247    & 0.863    & -2.257**  & 1.636***\\
                                         & (6.212)    & (1.657)    & (3.098)    & (0.760)    & (0.576)    & (0.986)    & (0.975)    & (0.885)    & (0.069)   \\[0.5em]

Other guns per 100,000 (Standardized)    & 1.568    & 0.559    & 1.321    & 0.128    & 0.375    & 0.199    & -0.073    & -0.055    & 0.325***\\
                                         & (1.763)    & (0.402)    & (1.044)    & (0.216)    & (0.211)    & (0.334)    & (0.224)    & (0.342)    & (0.071)   \\[0.5em]

PAL per 100,000 (Standardized)           & -1.100    & -0.498    & 0.355    & -1.526*   & 0.155    & -0.210    & 0.601*   & -1.137*** & 0.328***\\
                                         & (1.916)    & (1.297)    & (0.777)    & (0.802)    & (0.404)    & (0.209)    & (0.322)    & (0.252)    & (0.095)   \\[0.5em]

RPAL per 100,000 (Standardized)          & 0.360    & -1.002    & 1.917*   & -0.037    & 1.033    & -0.386    & -0.694    & 0.970    & -0.662** \\
                                         & (4.045)    & (2.109)    & (0.946)    & (1.285)    & (0.666)    & (0.708)    & (0.562)    & (0.786)    & (0.234)   \\[0.5em]

Province, CMA and Year FE                &  Yes    &  Yes    &  Yes    &  Yes    &  Yes    &  Yes    &  Yes    &  Yes    &  Yes   \\
%CMA FE                                  &  Yes    &  Yes    &  Yes    &  Yes    &  Yes    &  Yes    &  Yes    &  Yes    &  Yes   \\
%Year FE                                 &  Yes    &  Yes    &  Yes    &  Yes    &  Yes    &  Yes    &  Yes    &  Yes    &  Yes   \\
Demographic Covariates                   &  Yes    &  Yes    &  Yes    &  Yes    &  Yes    &  Yes    &  Yes    &  Yes    &  Yes   \\[0.5em]
Observations                             &  225    &  225    &  225    &  225    &  225    &  225    &  225    &  225    &  225   \\[0.5em]
Dependant Variable Mean                  & 5.653    & 1.798    & 1.548    & 1.373    & 1.702    & 1.62    & .679    & .749    & .193   \\
\\
\hline \hline
\end{tabular*}
\begin{scriptsize}
\begin{singlespace}
  Notes: All coefficients reported have been multiplied by 1000. Authors' calculations. %
  Years range from 2013 - 2019. %
  Each observation is year by CMA. %
  All equations are estimated using ordinary least squares. %
  Columns vary by their different types of crimes or violations. %
  Columns 1 is Criminal Code violation 150 (per 100,000) - total firearms violations, use of, discharge, pointing. %
  Column 2 is Criminal Code violation 1450 (per 100,000) - discharge firearms with intent. %
  Column 3 is Criminal Code violation 1455 (per 100,000) - using firearms in commission of offence. %
  Column 4 is Criminal Code violation 3390 (per 100,000) - firearms documentation or administrations. %
  Column 5 is Criminal Code violation 3395 (per 100,000) - unsafe storage of firearms. %  
  Column 6 is Criminal Code violation 110 (per 100,000) - homicide. %
  Columns 7 is Criminal Code violation 1110 (per 100,000) - murder, first degree. %
  Columns 8 is Criminal Code violation 1120 (per 100,000) - murder, second degree. %
  Columns 9 is Criminal Code violation 1130 (per 100,000) - manslaughter. % 
  These variables include: the average age of a person, the percentage of
  people that are female, the percentage of people that are married, the
  number of people receiving EI or Benefits, and the number of people
  receiving child benefits. %
  Standard errors are constructed using a CRVE and clustered 
  at the province-level.
  * significant at 10\%, ** significant at 5\%, *** significant at 1\%. 
\end{singlespace}
\end{scriptsize}
\end{table}

\end{landscape}

\setcounter{figure}{0}
\setcounter{table}{0}
\setcounter{section}{0}
\renewcommand{\thesection}{A\arabic{section}}
\renewcommand{\thetable}{A\arabic{table}}
\renewcommand{\thefigure}{A\arabic{figure}}

\newpage
\phantomsection
\stepcounter{section}
% Data Appendix 

\section{Data Appendix} \label{sec:apndx_data}

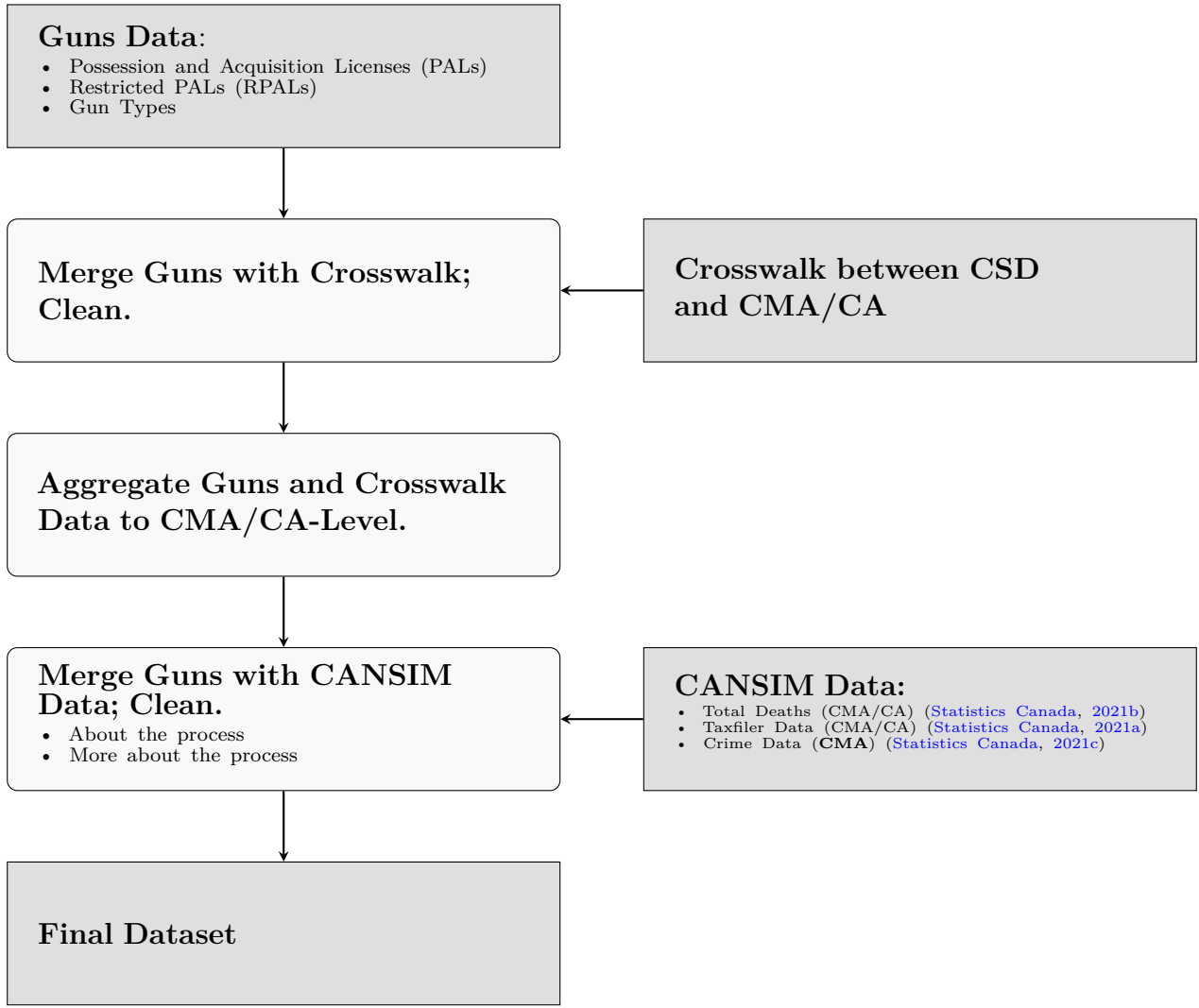
\begin{figure}[H] 
    \centering
    \caption{Data Preparation Workflow for Guns and Demographic Data}
    \label{fig:dataflow}

\begin{tikzpicture}[node distance = 1cm]

%%%%%%%%%%%%%%%%%%%%%%%%%%%%%%%%%%%%%%%%%%%%%%%%%%%%%%%%%%%%%%%%%%%%
% DRAW THE NODES 
%%%%%%%%%%%%%%%%%%%%%%%%%%%%%%%%%%%%%%%%%%%%%%%%%%%%%%%%%%%%%%%%%%%%

% GUNS DATA 
\node (guns) [dataset] {
	\parbox{\hsize}{
		\textbf{Guns Data}:%
		{%
		\vspace{-0.75em}
		\scriptsize%
		\begin{itemize}[leftmargin = * ]
			\setlength\itemsep{-0.75em}
			\item   Possession and Acquisition Licenses (PALs)
			\item   Restricted PALs (RPALs)
			\item   Gun Types 
		\end{itemize}
		\vspace{-0.75em}
		}
	}
};

% PROCESS OF MERGING AND CLEANING AND AGGREGATING
\node (mc1)[action, below of = guns, yshift = -2cm]{
	\parbox{\hsize}{
		\textbf{Merge Guns with Crosswalk; Clean.}
	}%
};%
%
% CROSSWALK 
\node (xwalk) [dataset, right of = mc1, xshift = \textwidth*0.46] {
	\parbox{0.75\hsize}{
		\textbf{Crosswalk between CSD and CMA/CA}
	}%
};%
%
% AGGREGATING FROM CSD TO CMA / CA 
\node (agg1)[action, below of = mc1, yshift = -2cm]{
	\parbox{\hsize}{
		\textbf{Aggregate Guns and Crosswalk Data to CMA/CA-Level.}
	}%
};%
%
% MERGE AND CLEAN 2
\node (mc2) [action, below of = agg1, yshift = -2cm] {
	\parbox{\hsize}{
		\textbf{Merge Guns with CANSIM Data; Clean.}
	{%
	\vspace{-0.75em}
	\scriptsize%
	\begin{itemize}[leftmargin = * ]
		\setlength\itemsep{-0.75em}
		\item   About the process 
		\item 	More about the process 
	\end{itemize}
		\vspace{-0.75em}
		}%
	}%
};%
%
% CANSIM DATA
\node (cansim) [dataset, right of = mc2, xshift = \textwidth*0.46] {
	\parbox{\hsize}{
		\textbf{CANSIM Data:}
		{%
		\vspace{-0.75em}
		\tiny%
		\begin{itemize}[leftmargin = * ]
			\setlength\itemsep{-0.75em}
			\item   Total Deaths (CMA/CA) \citep{StatCan_PopChange}  
			\item 	Taxfiler Data (CMA/CA) \citep{StatCan_TaxFilers}
			\item 	Crime Data (\textbf{CMA}) \citep{StatCan_Crime}
		\end{itemize}
		\vspace{-0.75em}
		}%
	}%
};%
%
% FINAL DATA 
\node (fin) [dataset, below of = mc2, yshift = -2cm] {\textbf{Final Dataset}};
%
%
%%%%%%%%%%%%%%%%%%%%%%%%%%%%%%%%%%%%%%%%%%%%%%%%%%%%%%%%%%%%%%%%%%%%
% DRAW THE ARROWS 
%%%%%%%%%%%%%%%%%%%%%%%%%%%%%%%%%%%%%%%%%%%%%%%%%%%%%%%%%%%%%%%%%%%%
\draw [arrow] (guns) -- (mc1); 
\draw [arrow] (xwalk) -- (mc1);
\draw [arrow] (mc1) -- (agg1);
\draw [arrow] (agg1) -- (mc2);
\draw [arrow] (cansim) -- (mc2);
\draw [arrow] (mc2) -- (fin);
\end{tikzpicture} %

\end{figure}

\begin{figure}[H]

\begin{center}
\caption{RCMP Application for Gun Licenses - Information Sheet}    
\label{apndx_infoSheet}   
\includegraphics[width=0.9\textwidth]{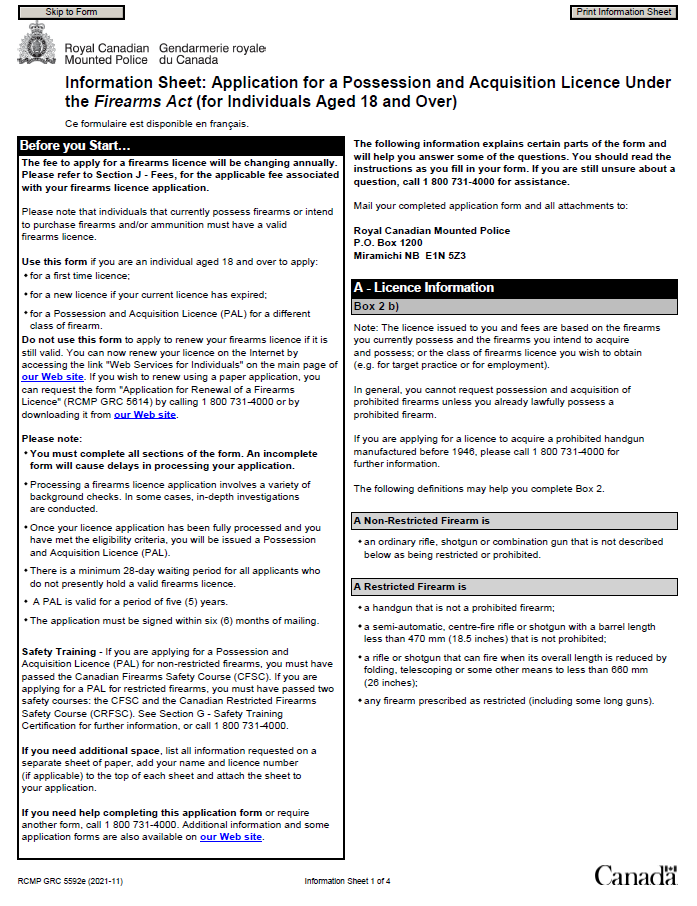}
\end{center}

\begin{scriptsize}
Application for a Possession and Acquisition Licence Under the 
\textit{Firearms Act} (for Individuals Aged 18 and Over). Retrieved
January 2022 from \url{https://www.rcmp-grc.gc.ca/en/firearms/firearms-forms}.
\end{scriptsize}

\end{figure}

The flowchart in Figure \ref{fig:dataflow} depicts the steps taken to clean
and merge the various data sets used in this analysis.

\subsection*{Cleaning}

\subsubsection*{Cleaning the Firearms Data}
Most cleaning associated with the firearms data amounted to making the
unit of observation suitable to be merged with the CSD to CMA/CA
crosswalk dataset. We imputed zeros for those CSDs which had missing
observations for a given year. For PALS and RPALS, three observations
were dropped in total: two because they were redundant and one
because it was empty. For the firearms type data, the process was
similar: clean it to make the unit of observation more suitable for
the merge with the crosswalk while imputing zeros for omitted 
values. 

\subsubsection*{Cleaning the CANSIM Data: Deaths, Taxfilers and Crimes}
Most cleaning done to these datasets was associated with tidying
up the observations in order to better match it to the crosswalk. 
Most of this amounted to string manipulation. In some specific 
instances, like the CMA of ``Greater Sudbury / Grand Sudbury'' we
had to rename some observations. Another example is the residual
group for provinces' ``non-CMA and/or non-CA'' which was named
inconsistently across the datasets. A final example is the changing
of a double dash to a single dash, or trimming the spaces between
dashes.  

\subsection*{Merging and Cleaning}

\subsection*{Merging the Firearms data with One another - PALS, RPALS, and Gun Types.}
The PAL and RPAL datasets merged perfectly with one another 
for all 3753 rows. However, merging the RPAL + PAL dataset
with the types of firearms dataset, our merge was successful for only
3320 rows, with 252 from the master dataset and 175 from the
``using'' dataset (the one being merged into the master dataset)
being unmatched. 128 rows were found to be 
merged unsuccessfully yet having discernibly the same identifying
variables and were corrected. Some observations were found to be 
redundant and therefore dropped. The final dataset resulted in 3581 
matched rows, with 125/252 from the master dataset and 135/175 of 
the using datasets being matched correctly. 70 rows were dropped 
since they were redundant.

This code concluded with reshaping the dataset from wide -- columns 
were year-by-statistic. The final number of observations was 25,067. 

\subsection*{Merging Firearm Types and Licenses Dataset with CSD to CMA/CA Crosswalk}
The firearms types and licenses dataset (all three merged, with their 
specific variables concatenated as columns) merged with the CSD 
to CMA/CA cross walk forming 23,499 / 25,067 successful matches. 
It should be noted that this is many-to-one merge since each 
geographic region corresponds to 7 rows (the number of years) of 
the firearms types and licenses dataset. There 79 CSDs which did not 
match in the crosswalk.

\newpage
\phantomsection
\stepcounter{section}
% RCMP License Application 

\section{RCMP Application for Firearms}  \label{sec:apndx_gunsApp}

\singlespacing

%%%%%%%%%%%%%%%%%%%%%%%%%%%%%%%%%%%%%%%%%%%%%%%%%%%%%%%%%%%%%%%%%%%
% INFORMATION SHEET 
%%%%%%%%%%%%%%%%%%%%%%%%%%%%%%%%%%%%%%%%%%%%%%%%%%%%%%%%%%%%%%%%%%%

\newpage

%%%%%%%%%%%%%%%%%%%%%%%%%%%%%%%%%%%%%%%%%%%%%%%%%%%%%%%%%%%%%%%%%%%
% PERSONAL HISTORY SHEET 
%%%%%%%%%%%%%%%%%%%%%%%%%%%%%%%%%%%%%%%%%%%%%%%%%%%%%%%%%%%%%%%%%%%
\begin{figure}[H]
\begin{center}
\caption{RCMP Application for Gun Licenses - Personal History}  
\label{apndx_personalHistory}      
\includegraphics[width=0.9\textwidth]{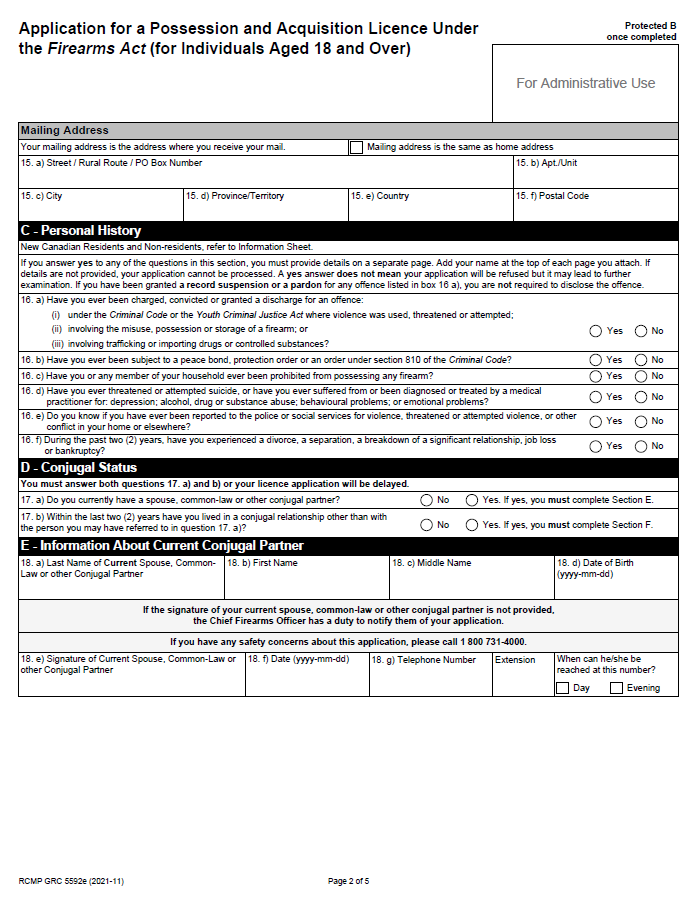}
\end{center}

\begin{scriptsize}
Application for a Possession and Acquisition Licence Under the 
\textit{Firearms Act} (for Individuals Aged 18 and Over). Retrieved
January 2022 from \url{https://www.rcmp-grc.gc.ca/en/firearms/firearms-forms}.
\end{scriptsize}

\end{figure}

\newpage
\phantomsection
\stepcounter{section}
%% CREATING THE FLOW CHART FOR THE CONCEPTUAL FRAMEWORK 

\begin{figure}

\begin{center}
\caption{Schematic Diagram for Firearms Markets and Uses \label{dia:arms_markets}}          
\end{center}
    
\begin{tikzpicture}[node distance = 1cm]

%%%%%%%%%%%%%%%%%%%%%%%%%%%%%%%%%%%%%%%%%%%%%%%%%%%%%%%%%%%%%%%%%%%%
% DRAW THE NODES 
%%%%%%%%%%%%%%%%%%%%%%%%%%%%%%%%%%%%%%%%%%%%%%%%%%%%%%%%%%%%%%%%%%%%
%
% Nature (Tier 1)
\node (Nature) [Nature] {\textit{Nature}};
%
%
% Types (Tier 2)
\node (Types) [Tier, below of = Nature, yshift = -4cm, xshift = -10cm] {\textit{Types} ($\theta$)};
\node (Never) [ConFrame, below of = Nature, yshift = -4cm, xshift = -4.5cm] {Never};
\node (Legal) [ConFrame, below of = Nature, yshift = -4cm, xshift = -1.5cm] {Legal};
\node (Either) [ConFrame, below of = Nature, yshift = -4cm, xshift = 1.5cm] {Either};
\node (Illicit) [ConFrame, below of = Nature, yshift = -4cm, xshift = 4.5cm] {Illicit};
%
%
% Markets (Tier 3))
\node (Markets) [Tier, below of = Types, yshift = -4cm, xshift = 0cm] {\textit{Markets} ($M$)};
\node (Formal) [ConFrame, below of = Types, yshift = -4cm, xshift = 10cm] {Formal};
\node (Informal) [ConFrame, below of = Types, yshift = -4cm, xshift = 13cm] {Informal};
%
%
% Uses (Tier 4)
\node (Uses) [Tier, below of = Markets, yshift = -4cm, xshift = 0cm] {\textit{Uses} ($\nu$)};
\node (Sport) [ConFrame, below of = Markets, yshift = -4cm, xshift = 8.5cm] {Sport};
\node (SelfHarm) [ConFrame, below of = Markets, yshift = -4cm, xshift = 11.5cm] {Self Harm};
\node (Crime) [ConFrame, below of = Markets, yshift = -4cm, xshift = 14.0cm] {Crime};

%
%
%%%%%%%%%%%%%%%%%%%%%%%%%%%%%%%%%%%%%%%%%%%%%%%%%%%%%%%%%%%%%%%%%%%%
% DRAW THE ARROWS 
%%%%%%%%%%%%%%%%%%%%%%%%%%%%%%%%%%%%%%%%%%%%%%%%%%%%%%%%%%%%%%%%%%%%
% Nature to Types 
\draw [arrow] (Nature) -- (Illicit);
\draw [arrow] (Nature) -- (Either);
\draw [arrow] (Nature) -- (Legal);
\draw [arrow] (Nature) -- (Never);
%
% Types to Markets 
\draw [arrow] (Legal) -- (Formal);
\draw [dashed, arrow] (Either) -- (Formal);
\draw [dashed, arrow] (Either) -- (Informal);
\draw [arrow] (Illicit) -- (Informal);
%
% Markets to Use 
\draw [dotted, arrow] (Informal) -- (Sport);
\draw [arrow] (Informal) -- (SelfHarm);
\draw [arrow] (Informal) -- (Crime);
\draw [arrow] (Formal) -- (Sport);
\draw [arrow] (Formal) -- (SelfHarm);
\draw [dotted, arrow] (Formal) -- (Crime);
\end{tikzpicture} %
\\
DESCRIPTION: \textit{Nature} assigns individual $i$ type $\theta$, who chooses to access firearms in market $M$ which then allows use $\nu$.
\end{figure}
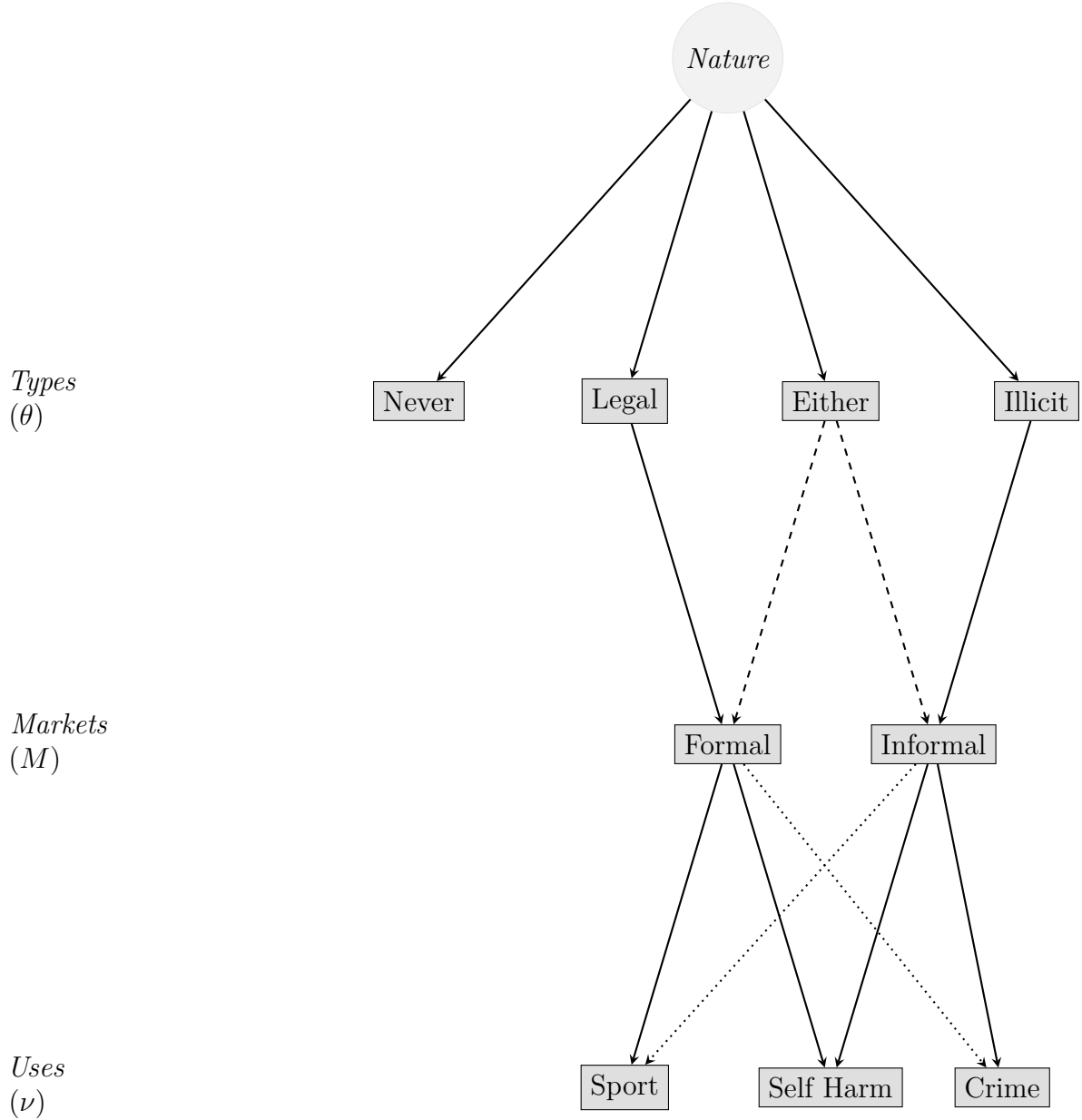

\clearpage

\begin{figure}[H]
	\begin{center}
    \caption{Directed Acyclic Graphs (DAGs) of Guns and Crime \label{fig:DAG}}
    \begin{subfigure}[b]{0.48\textwidth}
    \centering
			\begin{tikzpicture}[node distance = 1cm]
			%
			%%%%%%%%%%%%%%%%%%%%%%%%%%%%%%%%%%%%%%%%%%%%%%%%%%%%%%%%%%%%%%%%%%%%
			% nodes %
			%%%%%%%%%%%%%%%%%%%%%%%%%%%%%%%%%%%%%%%%%%%%%%%%%%%%%%%%%%%%%%%%%%%%
			\node (G) [xshift = 3.0cm] {\textit{Guns}};
			\node (C) [right of = G, xshift = 3.0cm] {\textit{Crimes}};
			\node (V) [draw, ellipse, dashed, above of = G, yshift = 1cm, xshift = 1.75cm] {\textit{Violence}};;
			%
			%%%%%%%%%%%%%%%%%%%%%%%%%%%%%%%%%%%%%%%%%%%%%%%%%%%%%%%%%%%%%%%%%%%%
			% edges %
			%%%%%%%%%%%%%%%%%%%%%%%%%%%%%%%%%%%%%%%%%%%%%%%%%%%%%%%%%%%%%%%%%%%%
			\draw [->, line width = 1] (G) -- (C);
			\draw[->, line width = 1] (V) -- (G);
			\draw[->, line width = 1] (V) -- (C);
			\end{tikzpicture}    
	\caption{Usual Concern with Guns and Crimes \label{DAG:usual}}      
    \end{subfigure}
    %
%    \\[1.0em]
    %
    \begin{subfigure}[b]{0.48\textwidth}
	\centering   
			\begin{tikzpicture}[node distance = 1cm]
			%
			%%%%%%%%%%%%%%%%%%%%%%%%%%%%%%%%%%%%%%%%%%%%%%%%%%%%%%%%%%%%%%%%%%%%
			% nodes %
			%%%%%%%%%%%%%%%%%%%%%%%%%%%%%%%%%%%%%%%%%%%%%%%%%%%%%%%%%%%%%%%%%%%%
			%
			\node (G) {\textit{Illegal Guns}};
			\node (C) [right of = G, xshift = 3.0cm] {\textit{Crimes}};
			\node (V) [draw, ellipse, dashed, above of = G, yshift = 1cm, xshift = 1.75cm] {\textit{Violence}};
			\node (L) [below of = G, yshift = -1.0cm] {\textit{Legal Guns}};
			%
			%%%%%%%%%%%%%%%%%%%%%%%%%%%%%%%%%%%%%%%%%%%%%%%%%%%%%%%%%%%%%%%%%%%%
			% edges %
			%%%%%%%%%%%%%%%%%%%%%%%%%%%%%%%%%%%%%%%%%%%%%%%%%%%%%%%%%%%%%%%%%%%%
			%
			\draw[->, line width = 1] (G) -- (C);
			\draw[->, line width = 1] (V) -- (G);
			\draw[->, line width = 1] (V) -- (C);
			\draw[->, line width = 1, red, dashed] (V) edge[bend right = 90] node[blue, rotate = 45] {{\Large X}} (L);
			\draw[->, line width = 1] (L) -- (C);
			\end{tikzpicture}  
	\caption{Legal versus Illegal Guns and Crimes \label{DAG:ours_crime}}       
    \end{subfigure}
    \end{center}
Notes: Proclivity for violence (\textit{Violence)} affects guns and crime but is hard to 
measure (as outlined by the dashed ellipse). This is depicted in subfigure \ref{DAG:usual}
and represents the usual concern with this estimation strategy. We disaggregate guns into 
those owned legally compared to those owned illegally in subfigure \ref{DAG:ours_crime}. 
Proclivity for violence (\textit{Violence}) affects \textit{illegal} guns and crime but is 
less likely to affect \textit{legal} gun use (as indicated in the figure by the red dashed 
edge with a blue `X' at its center). This is particularly true in the Canadian context which 
has strong firearms regulations relative to other contexts. The existing empirical literature
suggests a null relationship from legal firearms to crimes (\citet{Khalil2017} in the US as
well as a null relationship between (modest) volunteer firearms buybacks and 
crimes (\citet{Ferrazares2022} in the US).
\end{figure}
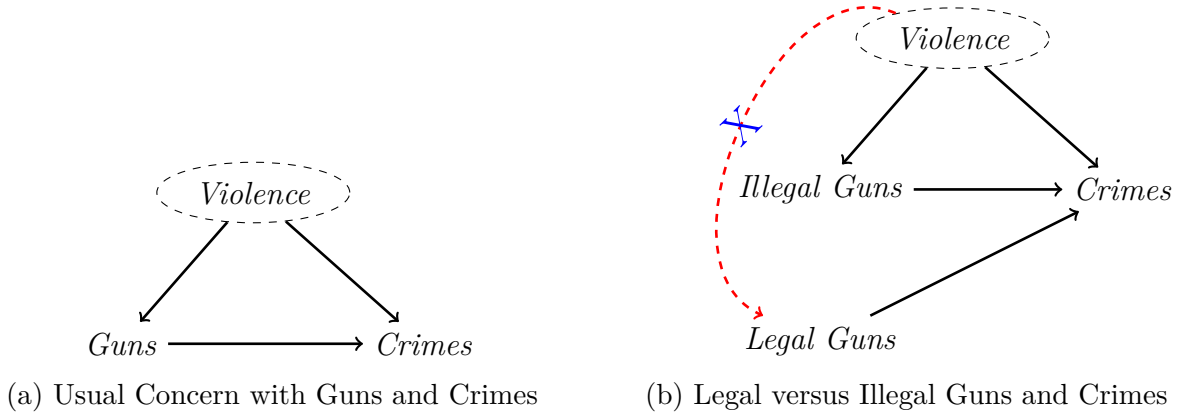

\vspace{1in}

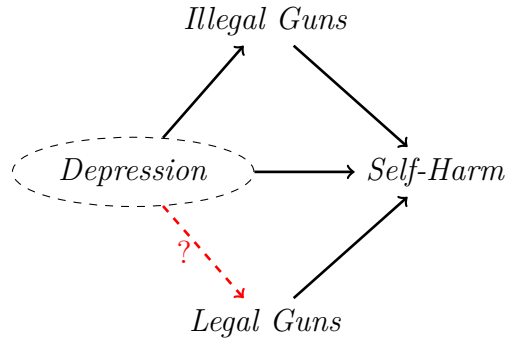
\begin{figure}[H]
\begin{center}
\caption{Legal versus Illegal Guns and Self-Harm DAG\label{Figure:DAG_selfharm}}          
\begin{tikzpicture}[node distance = 1cm]
%
%%%%%%%%%%%%%%%%%%%%%%%%%%%%%%%%%%%%%%%%%%%%%%%%%%%%%%%%%%%%%%%%%%%%
% nodes %
%%%%%%%%%%%%%%%%%%%%%%%%%%%%%%%%%%%%%%%%%%%%%%%%%%%%%%%%%%%%%%%%%%%%
%
\node (D) [draw, ellipse, dashed] {\textit{Depression}};
\node (SH)[right of = D, xshift = 3.0cm] {\textit{Self-Harm}};
\node (G) [above of = D, yshift = 1cm, xshift = 1.75cm] {\textit{Illegal Guns}};
\node (L) [below of = D, yshift = -1cm, xshift = 1.75cm] {\textit{Legal Guns}};
%
%%%%%%%%%%%%%%%%%%%%%%%%%%%%%%%%%%%%%%%%%%%%%%%%%%%%%%%%%%%%%%%%%%%%
% edges %
%%%%%%%%%%%%%%%%%%%%%%%%%%%%%%%%%%%%%%%%%%%%%%%%%%%%%%%%%%%%%%%%%%%%
%
\draw[->, line width = 1] (D) -- (SH);
\draw[->, line width = 1] (D) -- (G);
\draw[->, line width = 1] (G) -- (SH);
\draw[->, line width = 1, red, dashed] (D) edge node [left] {?} (L);
\draw[->, line width = 1] (L) -- (SH);
\end{tikzpicture}
\end{center}
Notes: Depression is likely to affect both \textit{illegal} gun users and \textit{legal} gun 
users. Though the indirect path from depression to self-harm through legal guns is expected 
to be mitigated but not eliminated by existing firearms regulation (as suggested by the red 
and dashed edge with a question mark over the edge's center). Specifically, a \textit{truthfully 
reported} history of suicidal thoughts makes obtaining a license all but impossible.  
Additionally, the `reference check' part of the licencing process specifically tells the 
references ``\textit{If you have any safety concerns about this application, please call 
1-800-731-4000.}'' Previous literature suggests a positive relationship between the dashed, 
red line from depression to legal firearms \cite{Perlis2022}. Previous literature suggests 
either a positive relationship between black arrow running from legal firearms to self-harm 
(as in \citet{Leigh2010} in Australia) or a null relationship between (modest) volunteer 
firearms buybacks and self harm (as in \citet{Ferrazares2022} in the US). 
\end{figure}

\end{document}